\documentclass[11pt]{article}

\usepackage{amsmath}
\usepackage{graphicx}
\usepackage{indentfirst}
\usepackage{amssymb}
\usepackage{cite}
\usepackage{color}
\usepackage{subfigure}

\begin{document}

\title{  The feature of shadow images and observed luminosity of the Bardeen black hole surrounded by different accretions  }

\date{}
\maketitle

\begin{center}
\author{Ke-Jian He,}$^{1}$\footnote{ kjhe94@163.com}
\author{Sen Guo,}$^{2}$\footnote{ sguophys@126.com}
\author{Shuang-Cheng Tan,}$^{1}$\footnote{shuangchengtan@cqu.edu.cn}
\author{Guo-Ping Li}$^{*}$\footnote{*corresponding author: gpliphys@yeah.net}\\

\vskip 0.25in
$^{1}$\it{Department of Physics and Chongqing Key Laboratory for Strongly Coupled Physics, Chongqing University, Chongqing 401331, China}\\
$^{2}$\it{ Guangxi Key Laboratory for Relativistic Astrophysics, School of Physical Science and Technology, Guangxi University, Nanning 530004, China}\\
$^{*}$\it{Physics and Space College, China West Normal University, Nanchong 637000, China}

%\affiliation[*]{corresponding author: gpliphys@yeah.net}
\end{center}
\vskip 0.6in
{\abstract
{In this paper, by exploring the photon motion in the region near the Bardeen black hole,  the shadow and observation properties of the black hole surrounded by various accretion models are studied. We analyzed the changes in shadow imaging and observation luminosity when the relevant physical parameters are changed. For the different spherical accretions background, one can find that the radius of shadow and the position of  photon sphere do not change, but the observation intensity of  shadow  in the infalling accretion model is significantly lower than that of the static case. When the black hole is surrounded by an optically and thin disk accretion,  the contribution of the photon rings, lensing rings and direct emission to the total observed flux has also been studied.  Under the  different forms of the emission modes, the result shows that the observed brightness  is mainly determined by direct emission,  while the lensing rings will provide a small part of the observation flux and the photon ring can provide a negligible observation flux. By comparing our results with the Schwarzschild spacetime, it is found that the existence or change of relevant status parameters  will greatly affect the shape and observation intensity of black hole shadow. These results support that the change of state parameter will affect the spacetime structure,  thus affecting  the observation feature of black hole shadows. }}

\thispagestyle{empty}
\newpage
\setcounter{page}{1}

\section{Introduction}\label{sec:intro}
Since the concept of black hole was put forward, people have been trying to find this mysterious physical object in the universe. Recently, the first image of supermassive black holes at the center of M$87^*$ galaxy, which was captured by the Event Horizon Telescope Collaboration (EHT), is a strong proof of the existence of the  black hole\cite{Akiyama:2019cqa,Akiyama:2019brx,Akiyama:2019sww,Akiyama:2019bqs,Akiyama:2019fyp,Akiyama:2019eap}. One can  observed that a bright ring appears in the background of dark area in the image, where the dark area in the center is called the shadow of  black hole and the bright area is photon sphere. It is commonly known that the strong gravity of  black hole deflects the light and forms the shadow of  black hole, which was also known as gravitational lensing effect\cite{Wal}. With the deepening of the study of black hole,  people pay more attention to the observation of black hole shadow, because it will bring new insights and development to the study of black hole\cite{Jaroszynski:1997bw,Shaikh:2018lcc,Gralla:2019drh,Allahyari:2019jqz,Li:2020drn,Banerjee:2019nnj,Vagnozzi:2019apd,Vagnozzi:2020quf,Safarzadeh:2019imq,Davoudiasl:2019nlo,Roy:2019esk,Roy:2020dyy,
Chen:2019fsq,Konoplya:2020bxa,Islam:2020xmy,Jin:2020emq,Guo:2020zmf,Wei:2020ght,Grenzebach:2014fha,Li:2021btf,Cunha:2019hzj, Amarilla:2011fx,Hou:2021okc,Zhong:2021mty,Hu:2020usx,Long:2019nox,Wang:2021art,Wang:2021ara,Long:2020wqj,Guerrero:2021pxt,Jafarzade:2020ova,Chang:2020lmg,Dey:2020bgo}.

In our universe, the real astrophysical black holes in galaxies are surrounded a large amount of high energy radiation material, which makes it possible to indirectly observe  the  black hole\cite{Porth:2019wxk}. The accretion flow around a black hole is usually not spherically symmetric, but the simplified ideal spherical model can provide strong support for the characteristics of black hole observation. It is generally accepted that the shape of the black hole shadow is standard circular geometry in the static spherically symmetric spacetime\cite{Synge:1966okc}. In \cite{Bar}, Bardeen pointed out that the  radius of shadow for Schwarzschild black hole  is $r_s=3 M$, where $M$ is the mass of black hole. Meanwhile, Bardeen also mentioned that the shadow shape of rotating black hole is closely related to angular momentum. In 1979, the simulated image of a black hole is shown for the first time,  in which the black hole surrounded by optically thin disk was considered  \cite{Luminet:1979nyg}. Taking into account the model in which the black hole surrounded by the optically thin spherical accretion, it is found that the  shadow  is  geometric characteristics of spacetime, and has nothing to do with the details of the accretion process\cite{Narayan:2019imo}. However,  the observed luminosity is affected by the  accretion flow model \cite{Falcke:1999pj,Bambi:2013nla,Zeng:2020dco,Guo:2021bwr,Qin:2020xzu}. When an optically and geometrically thin disk accretion is located on the equatorial plane of  black hole,  the observational peculiarity of  black hole shadow observed by the distant observer have been studied in\cite{Gralla:2019xty}. Gralla et al. classified the trajectories of light rays near the black hole,  whereby divided the rings  outside  the shadow area  into  the direct, photon ring, and lensing ring, and proposed that the observation intensity is dominated by the direct case. Furthermore, they also found that the size of the black center region depends on the specific emission model of accretion flow. Using the different accretion flow models, there is a series interesting observable features about the shadow and rings of the black hole in other gravitational spacetime background, which could refer \cite{Zeng:2020vsj, Peng:2020wun,Peng:2021osd, He:2022yse,Li:2021ypw,Zeng:2021mok,Zeng:2021dlj,Guerrero:2021ues,Gan:2021xdl,Li:2021riw,Guo:2021bhr}.

Most often, the black hole  has a singularity inside the horizon. While in \cite{bardeen1968non}, the model of a black hole with regular non-singular geometry  was proposed by Bardeen. In this system, an energy-momentum tensor is introduced, which is interpreted as the gravitational field of a nonlinear magnetic monopole charge. Since then, the physical properties of  Bardeen black hole has aroused people's interest, and  involve a wide range of quasinormal modes, energy distribution, radiation, thermodynamic behavior and so on \cite{Man:2013hpa,ayon2000bardeen,ayon1998regular,hayward2006formation,berej2006regular,Borde:1994ai,
Borde:1996df,Eiroa:2010wm,Pradhan:2014oaa,Hu:2020biz,Ghaffarnejad:2014zva,Saleh:2017vui,V.:2019ful,Narzilloev:2020qtd}. Indeed, the study of black hole shadow can explore the basic physical properties of spacetime.
Therefore, it is necessary to study the  shadow and observation characteristics in the context of Bardeen spacetime. In this paper, we focus on the shadow and observational luminosity of the Bardeen black hole surrounded by various accretion models, i.e., the static,  infalling spherical accretion and the optically thin disk accretion, which were regarded as the only background light source. By comparing the results in this work with  Schwarzschild spacetime, whereby we can study the influence of magnetic monopole charge on the shadow  and observation characteristics, due to the  charge parameter is  a remarkable role of Bardeen spacetime. Therefore, it may provide  a feasible method to distinguish the Bardeen spacetime from Schwarzschild spacetime.

The paper is organized as follows. In section 2, we have studied the associated  trajectories of the light ray near the Bardeen black hole, as well as the radius of photon sphere and the shadow of  black hole when the state parameters changed. In section 3, we analyzed  the shadow image and luminosity of the black hole surrounded by the different  spherical accretion models. In section 4, we considered thin disk emission near the black hole, and compared the observational appearance with the different emission profiles. Finally, we discussed our results and conclusions in section 5. In this paper, we use the units $G  = c = 1$.

\section{Null geodesic in the  Bardeen  space-time }
\label{light}
We would like to consider the spherically symmetric  spacetime, and the  metric of Bardeen regular black hole  as follows\cite{bardeen1968non},
\begin{equation}
d s^2=-A(r) dt^2 +B (r)dr^2 +C (r) \left({d\theta }^2+ \sin ^2 \theta {d \varphi }^2 \right),\label{EQ2.1}
\end{equation}
with
\begin{equation}
A(r)=1-\frac{2 M r^2}{\left(g^2+r^2\right)^{3/2}},\quad B(r)=\frac{1}{A(r)}, \quad C (r)=r^2. \label{EQ2.2}
\end{equation}
In which, $M$ is the mass of the black hole, and $g$ is the monopole charge of a self-gravitating magnetic field described by a nonlinear electrodynamics source. Noted that the spacetime structure could recover to the Schwarzschild spacetime in the case of $g=0$. As mentioned in \cite{Man:2013hpa}, the metric function  can be approximated as
\begin{equation}
A (r)\sim 1-\frac{2 M}{r}+\frac{3 g^2 M}{r^3}+{\mathcal{O}}(\frac{1}{r^5}).\label{EQ2.3}
\end{equation}
It is obvious that the form of the above equation is different from the $Reissner-Nordstr\ddot{o}m$ spacetime.  And, the horizon of the Bardeen black hole can be obtain by
\begin{equation}
A (r)=0, \label{EH}
\end{equation}
where the largest positive root is the event horizon of the black hole. It is worth noting that the monopole charge $g$ should follow the condition $g<\frac{4\surd3}{9}M$,  which is important  threshold for the existence of event horizon. For the case $g>\frac{4\surd3}{9}M$, there is no horizon and  we do not consider in this paper.
When the light ray passes from the vicinity of a massive object it is deflected due to the interaction between light and gravitational field of the massive object.  In doing so, we can use the  Euler-Lagrange equation to investigate the  motion behaviors of photons around the Bardeen black hole, which reads as
\begin{equation}
\frac{d}{d\lambda }\left(\frac{\partial \mathcal{L}}{\partial \dot{x}^{\mu }}\right)=\frac{\partial \mathcal{L}}{\partial x^{\mu }},\label{EQ2.5}
\end{equation}
where $\lambda$ and $\dot{x}^{\mu }$ are the affine parameter and four-velocity of the photon, respectively. For the metric function (\ref{EQ2.1}),  the Lagrangian $\mathcal{L}$ can be expressed as
\begin{equation}
\mathcal{L}=\frac{1}{2} \mathit{g}_{\mu \nu } \dot{x}^{\mu } \dot{x}^{\nu }=\frac{1}{2} \left(-A (r) \dot{t}^2+B (r) \dot{r}^2+C (r) \left(\dot{\theta} ^2+  \sin\theta ^2 \varphi ^2\right)\right).\label{EQ2.6}
\end{equation}
In the spherically symmetric spacetime,  we discuss the motion of photons confined to the equatorial plane of a black hole\cite{Synge:1966okc,Bardeen:1972fi,Gralla:2019drh}, and  have the restrictive conditions  $\theta =\frac{\pi }{2}$ and $\dot{\theta} =0$.
Due to  the $t$ and $\theta$ coordinates can not explicitly determine the coefficient of the metric equation, there are two conserved quantities $E$ and $L$, representing energy and angular momentum, respectively. That is
\begin{align}
E=-\frac{\partial \mathcal{S}}{\partial \dot{t}}=A (r) \dot{t},\quad
L=\frac{\partial \mathcal{S}}{\partial \dot{\varphi} }=r^2 \dot{\varphi}.\label{EQ2.7}
\end{align}
For null geodesic $\mathit{g}_{\mu \nu } \dot{x}^{\mu } \dot{x}^{\nu }=0$, we can obtain the orbit equation
\begin{equation}
\dot{r}^2+V_{eff}=\frac{1}{b^2}. \label{EQ2.8}
\end{equation}
In which, $V_{eff}$ is the effective potential, which is given by
\begin{equation}
V_{eff}=\frac{1}{r^2} (1-\frac{2 M r^2}{\left(g^2+r^2\right)^{3/2}}),\label{EQ2.9}
\end{equation}
Moreover, $b$ is called the impact parameter, which is the ratio of angular momentum to energy, i.e., $b = J/E=\frac{r^2 \dot{\varphi}}{A (r) \dot{t}}$. In the spacetime (\ref{EQ2.1}), there exist a critical impact curve with radius $b_p$,
and  the light from this curve will asymptotically approach a bound photon orbit. Indeed, the  shadow is represented the interior of the critical curve, and the surface of bound  is the photon sphere.
In the position of the photon sphere, the effective potential should follow the condition $\dot{r}=0$ and $\ddot{r}=0$, which means\cite{Bambi:2013nla}
\begin{align}
V_{eff}=\frac{1}{b ^2},\quad 2 V'_{eff}=0.\label{VE}
\end{align}
From Eq.(\ref{VE}), one can find the relation between  the photon  sphere radius $r_p$ and critical impact parameter $b_p$ is contented
\begin{align}
r_p{}^2-b_p^2 A(r)=0,\quad 2 b_p^2 A(r)^2-r_p{}^3 A'(r)=0.\label{EQ2.10}
\end{align}
For the distant observer ($r_o\rightarrow\infty$), $b_p$ is the value of  shadow radius that can be seen. For the different values of the  monopole charge $g$, we can get the different numerical results of the  photon  sphere radius $r_p$,  the shadow radius $b_p$ and the event horizon $r_h$ where $M = 1$, and listed in  Table 1.
\begin{center}
{\footnotesize{\bf Table 1.} The value of  relevant physical quantities under the  different values of  monopole charge $g$, in which $M = 1$.\\
\vspace{1mm}
\begin{tabular}{ccccccccc}
\hline       &{$g=0.1$}      & {$g=0.2$}     & {$g=0.3$} &   {$g=0.4$} &     {$g=0.475$} & {$g=0.6$} & {$g=0.7$}  & {$g=0.75$}\\    \hline
{$r_{p}$} & {2.99164}        &{2.96617}         &{2.9224}     &{2.85798}        &{2.7937}  &{2.64674}   &{2.47495}  &{2.35745}                  \\
{$b_{p}$} & {5.18747}         &{5.16108}       &{5.11594}     &{5.05508}        &{4.98508}    &{4.83975}   &{4.67719}   &{4.57212}                              \\
{$r_{h}$}  & {1.99247}        &{1.96946}       &{1.92962}      &{1.87022}     &{1.80981}   &{1.66546}    &{1.47436}   &{1.29904}                              \\
\hline
\end{tabular}}
\end{center}
From Table 1, it can be find that all the related physical quantities  $r_p$,   $b_p$ and  $r_h$  show  a decreasing trend with  the increase of monopole charge $g$. Therefore, the appearance of  monopole charge $g$ will lead to the reduction of the black hole area compared with the Schwarzschild black hole. With the help of introducing parameter $u_0=1/r$, we can rewrite the motion equation of photons, and use it to depict the trajectory of light ray, that is
\begin{align}
 \mathcal{V}(u_0)=\frac{du_0}{d\varphi }=\sqrt{\frac{1}{b^2}-u_0^2 \left(1-\frac{2 M}{u_0^2 \left(g^2+\frac{1}{u_0^2}\right){}^{3/2}}\right)}.\label{EQ2.11}
\end{align}
It is worth noting that the geometry of geodesics depends entirely on the Eq.(\ref{EQ2.11}), which as a function of the impact parameter $b$.  In the case of $b<b_p$, the light ray is always trapped by the black hole and cannot reach infinity distance. In the case of $b>b_p$, the light ray  will be deflected at the radial position $u_0$ where $\mathcal{V}(u_0)=0$, and then move away from the black hole reach infinity distance. When $b=b_p$, the photons are in a state of rotation around the black hole, neither falling into the black hole nor falling into the black hole. In the following sections, we further study the shadow of black hole surrounded by  various profiles of accretion flow.

%%%%%%%%%%%%%%%%%%%%%

\section{Image of the Bardeen black hole with spherical accretion
 }
With the help of equation (\ref{EQ2.11}), on can find that  the trajectory of a light ray  changed as the change of the value of monopole charge $g$ and impact parameter $b$. Therefore, we simply describe the  path of light rays under the relevant physical parameter, as shown in Figure 1.
\begin{figure}[h]
\centering % \begin{center}/\end{center} takes some additional vertical space
\subfigure[$g=0.1$]{\includegraphics[scale=0.35]{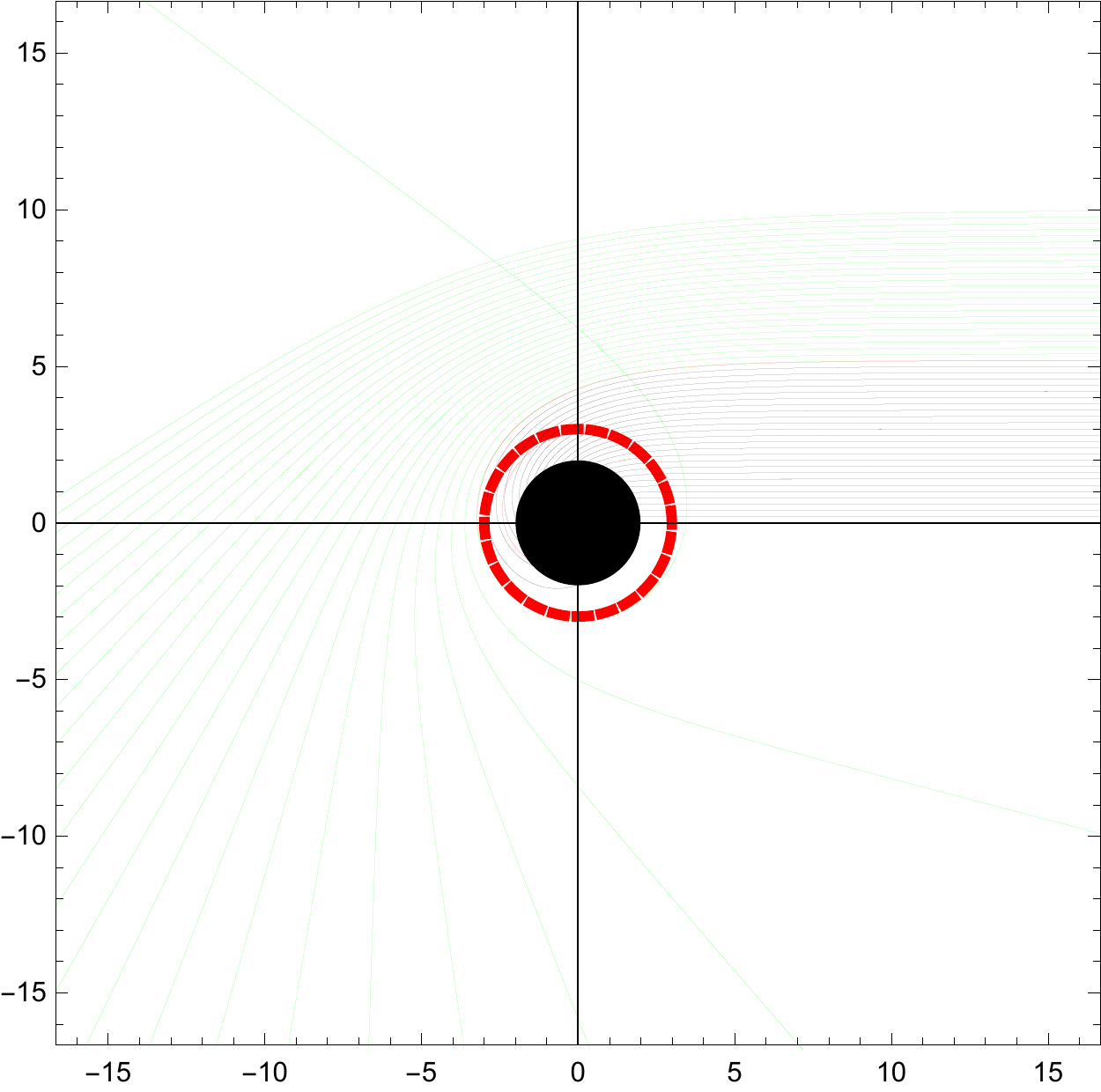}}
\subfigure[$g=0.475$]{\includegraphics[scale=0.35]{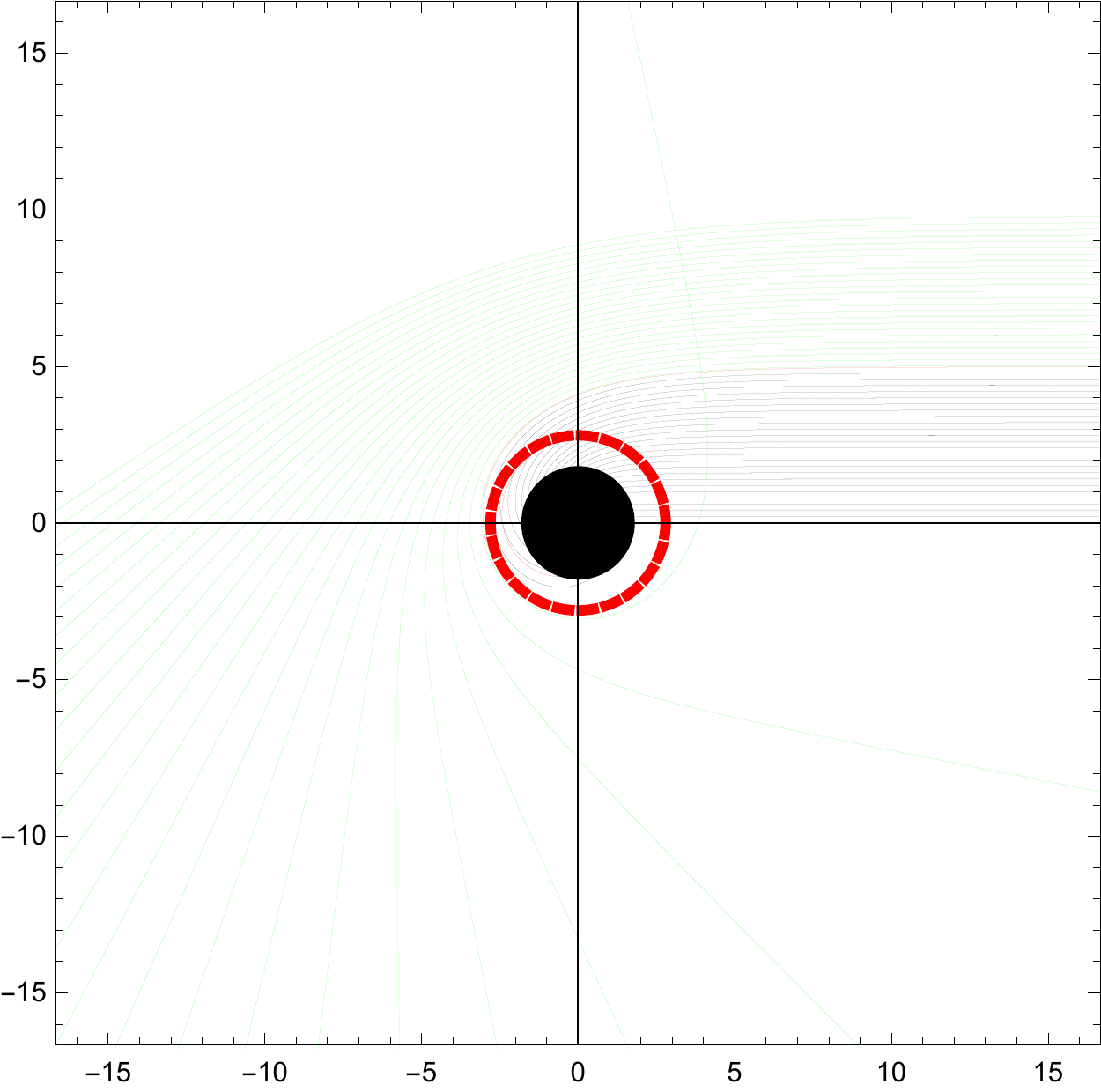}}
\subfigure[$g=0.75$]{\includegraphics[scale=0.35]{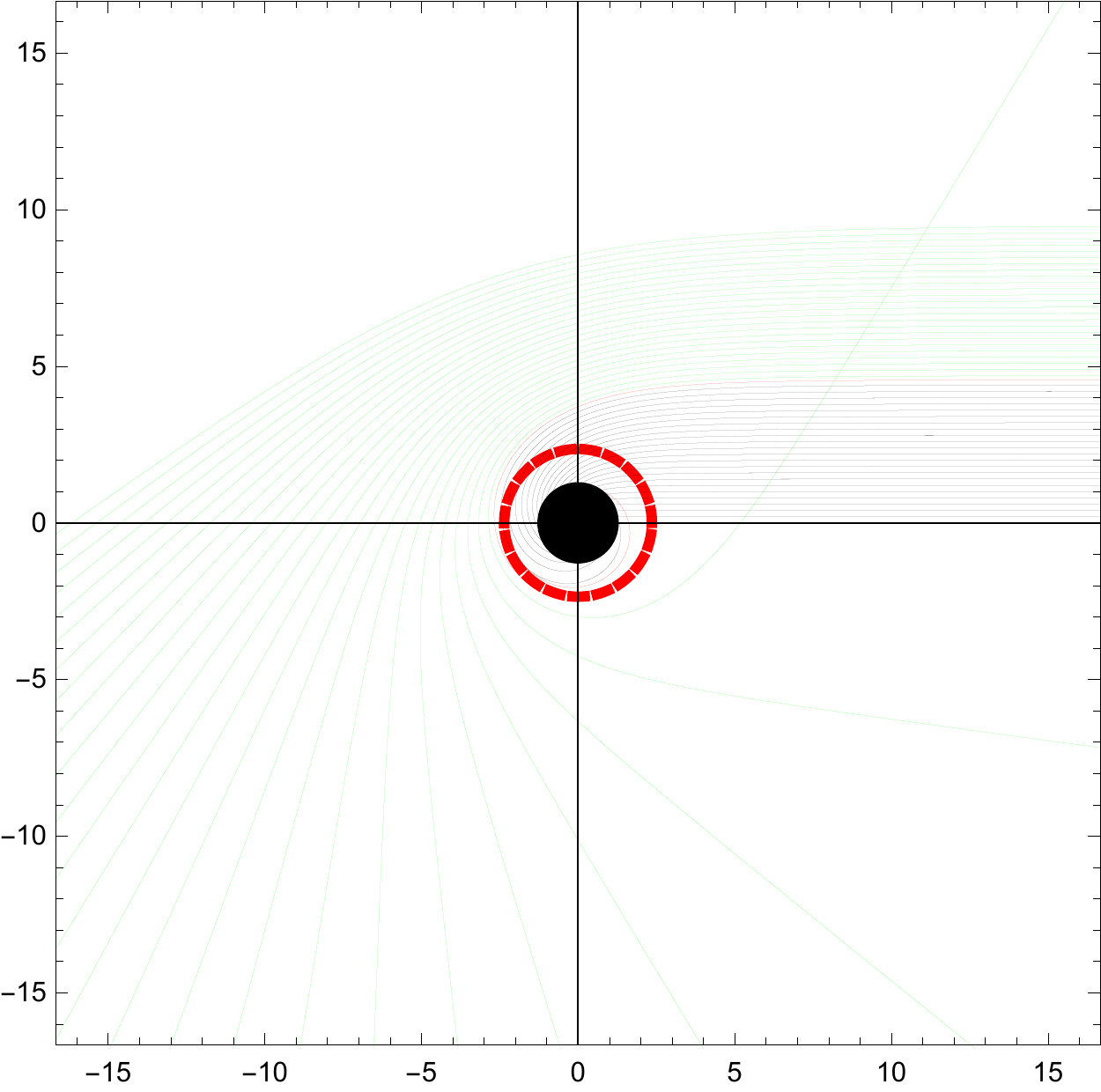}}
\caption{\label{fig1} The  trajectory of the light ray for the  different values of monopole charge $g$  in the polar coordinates $(r, \varphi)$ with $M=1$. In which, the black hole is shown as a black disk, and the red dotted line represents the position of the photon sphere.  }
\end{figure}

In Figure 1,  the red lines, green lines and black lines  correspond to $b=b_p$, $b>b_p$ and $b<b_p$, respectively. When the value of monopole charge $g$  increases,  it is obvious that the the radius of the photon sphere $r_p$ decreases, as well as the event horizon $r_h$. On the other hand, the deflection rate of light ray is different under the different  values of monopole charge $g$. Namely, the light density obtained by distance observers is different, which naturally leads to the different observation intensity casted by black hole shadow. For the distance observer, we want to  study the observation luminosity of the black hole shadow, in which the  accretion flow around the black hole is the only light source. Then,  two simple relativistic spherical accretion models is  introduced, that is,  the static spherical accretion and the infalling spherical accretion.

\subsection{ The shadow and photon sphere with static spherical accretion flow}

In the background of the black hole being wrapped by static spherical accretion flow, we can integrate the specific emissivity along the photon path $\gamma_i$, so as get the specific intensity $\mathcal{I}_{{obs}}$
for the distance observer  (usually measured in $\rm erg s^{-1} cm^{-2} str^{-1} Hz^{-1}$). That is \cite{Jaroszynski:1997bw, Bambi:2013nla}
\begin{equation}
 \mathcal{I}_{{obs}}(\nu _0)=\int _{\gamma _i}{g_s}^3 {j} (\nu _e) dl_{{prop}}, \label{EQ3.1}
\end{equation}
and
\begin{equation}
g_s=\frac{\nu _o}{\nu _e}. \label{EQ3.2}
\end{equation}
It is worth mentioning that we focus on the case that the emitter is fixed in the rest frame. In the above equation, $g_s$ is the redshift factor and $\nu _e$ is  the photon frequency at the emitter. In addition,  ${j} (\nu _e)$ is the emissivity per-unit volume at the emitter,  $dl_{{prop}}$ is the infinitesimal proper length, and  $\nu _0$ is the observed photon frequency, respectively. In the spacetime(\ref{EQ2.1}), the redshift factor $g_s=A(r)^{1/2}$. For the specific emissivity ${j} (\nu _e)$, we consider that the emission is monochromatic with  the emission frequency $\nu _f$  and the emission radial profile is $1/r^2$,  that is,
\begin{equation}
{j} (\nu _e)\sim\frac{\delta  \left(\nu _e-\nu _f\right)}{r^2}, \label{EQ3.3}
\end{equation}
where $\delta$ is the  delta function. In addition, we can obtain the proper length  $dl_{{prop}}$  is
\begin{equation}
dl_{{prop}}= \sqrt{\frac{1}{A (r)}+r^2 \left(\frac{d\varphi }{dr}\right)^2} dr.\label{EQ3.4}
\end{equation}
Then, the expression of the observed specific intensity  can be rewritten as
\begin{equation}
\mathcal{I}_{{obs}}(\nu _o )=\int _{\gamma _i}\frac{(A (r))^{3/2}}{r^2}  \sqrt{\frac{1}{A (r)}+r^2 \left(\frac{d\varphi }{dr}\right)^2} dr.\label{EQ3.5}
\end{equation}
In Eq.(\ref{EQ3.5}), the total  observed  intensity $ \mathcal{I}_{{obs}}(\nu _o )$ as a function of impact parameter $b$, which means that the change of relevant physical quantity will directly affect the  specific intensity observed by  a distant observer. We take $g=0$, $g=0.475$ and $g=0.75$ as examples, and the  change trend of  observed intensity with the change $b$ is given in  Figure 2.
\begin{figure}[h]
\centering % \begin{center}/\end{center} takes some additional vertical space
\includegraphics[width=0.50\textwidth]{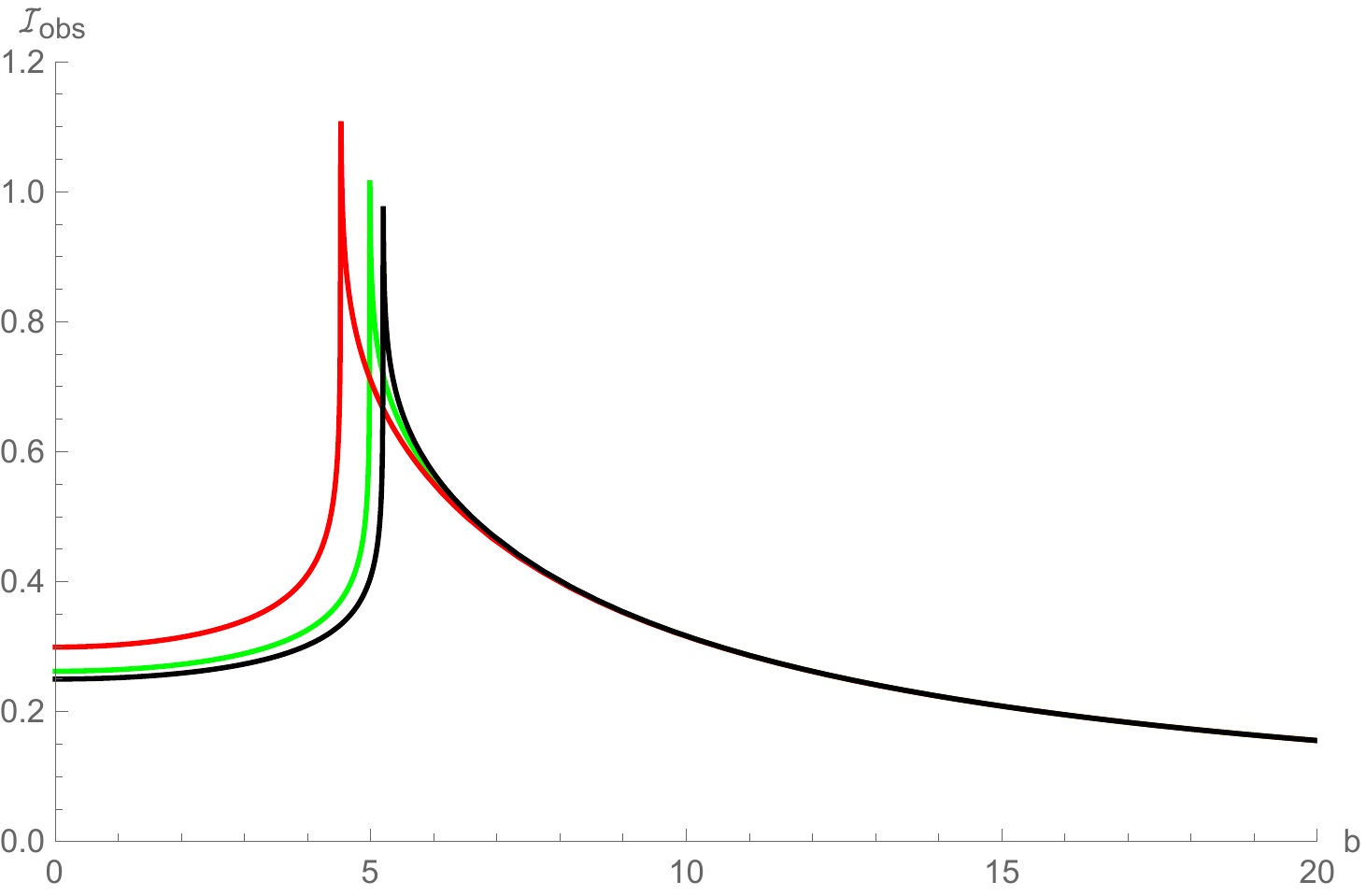}
% "\includegraphics" is very powerful; the graphicx package is already loaded
\caption{\label{fig2}  The observed specific intensity  of the black hole is surrounded by  the static spherical accretion flow, in which the black line, green line and red line correspond to $g=0$, $g=0.475$ and $g=0.75$. Here, we take M = 1.}
\end{figure}

It is clear that  the observation intensity increases slowly with the increase of $b$ ($b<b_p$), but increases rapidly near the photon sphere ($b=b_p$) until  reaches to the peak at the position of the photon sphere. After that, the observation intensity decreases with the increase of $b$ ($b>b_p$) and  at the infinity $\mathcal{I}_{{obs}}(\nu _o )\rightarrow0$. In addition, the larger value of $b$, the stronger  specific intensity observed. For the case $g =0$( the Schwarzschild spacetime), the peak of observation intensity is obviously smaller than that when $g=0.45$ ( or $g=0.75$). In other words, the observational intensity of the Bardeen black hole is stronger than that of the Schwarzschild spacetime, and the appearance of magnetic monopole charge $g$ increases the light density, so that the  observer can obtain more luminosity. The two-dimensional intensity map in celestial coordinates casted by the black hole  is shown in  Figure 3.
\begin{figure}[h]
\centering % \begin{center}/\end{center} takes some additional vertical space
\includegraphics[width=.325\textwidth]{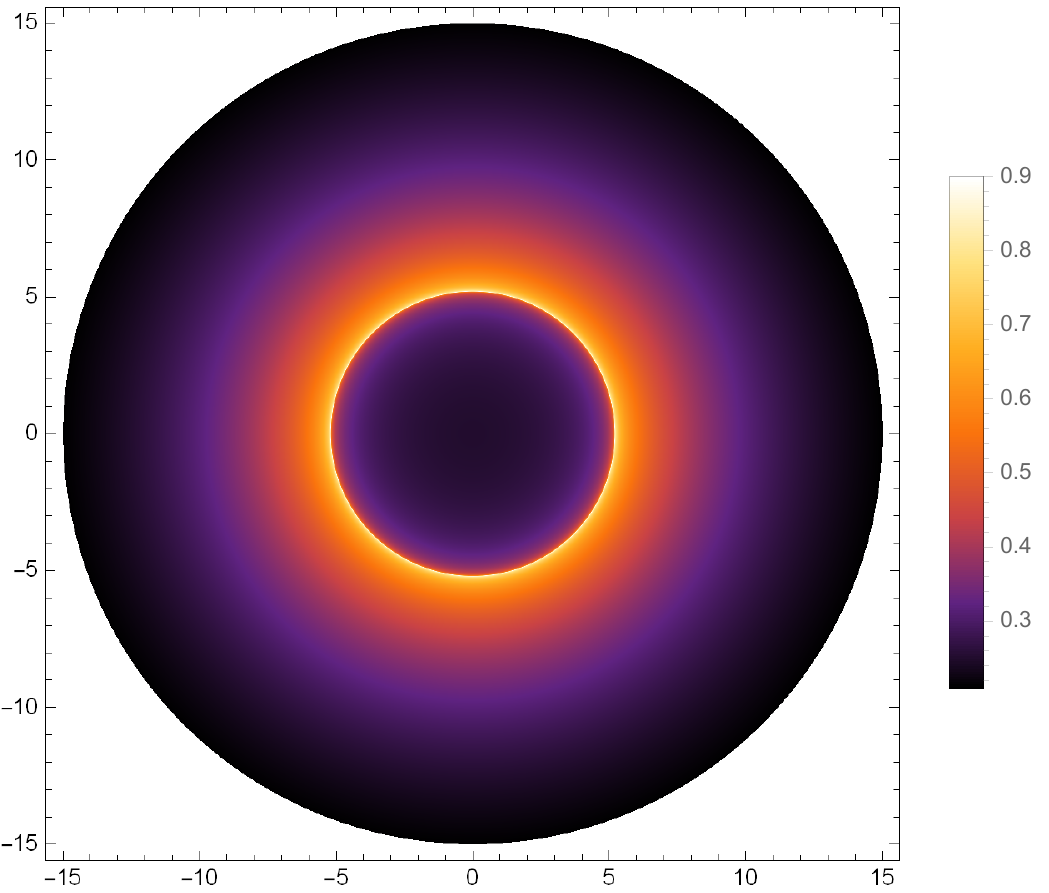}
\hfill
\includegraphics[width=.325\textwidth]{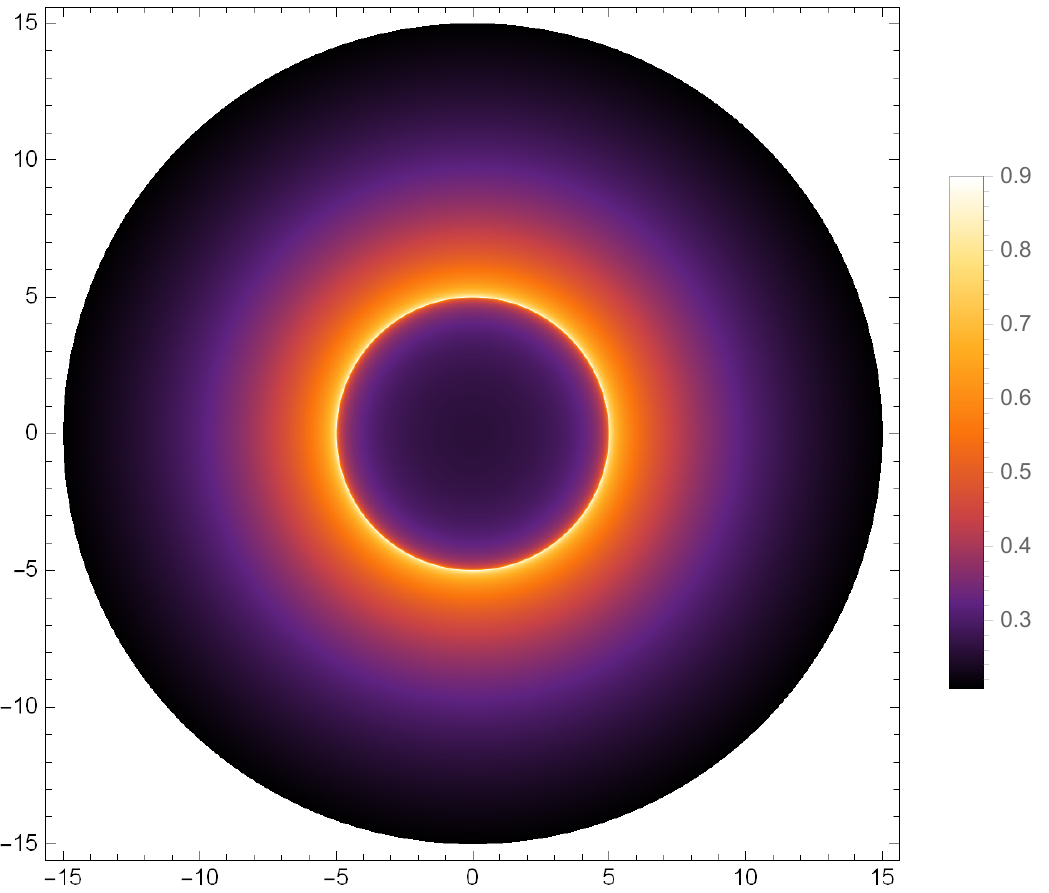}
\hfill
\includegraphics[width=.325\textwidth]{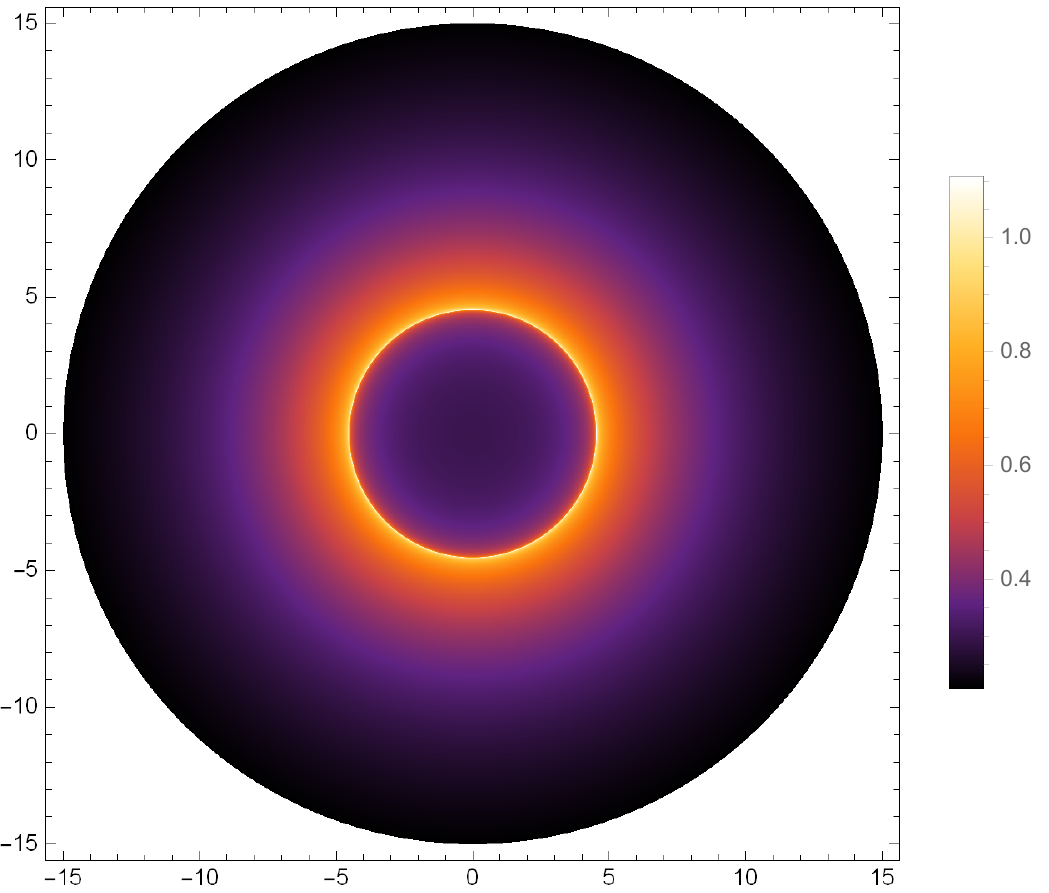}
\caption{\label{fig3}  Image of the black hole shadow with the static spherical accretion for the different values of monopole charge $g$, in which $g=0$ (left panel),$g=0.475$(mid panel) and $g=0.75$ (right panel).  }
\end{figure}

From Figure 3, there exist a bright ring with the strongest luminosity, which is  the position of  photon sphere. Due to the tiny fraction of the radiation  can escape from the black hole, the inner region of the photon sphere is not completely black, and there is small observation luminosity near the photon sphere. Obviously, the maximum  luminosity of the shadow image  enhanced with the increase of monopole charge $g$, while the size of shadow and the radius of photon sphere  decreased.

\subsection{The shadow and photon sphere with infalling spherical accretion flow}
In the background of  black hole surrounded by the infalling spherical accretion,  the radiating gas moves towards the black hole along the radial direction. Indeed, the infalling spherical accretion is closer to the real situation than the static spherical accretion model. We need to pay attention to that the equation (\ref{EQ3.5}) is still suitable in the infalling  accretion model, but the redshift factor is different from the static case. In the infalling accretion model, the redshift factor  is related to the velocity of the accretion flow, namely
\begin{equation}
g_i=\frac{\mathcal{K}_{\rho } u _0^{\rho }}{\mathcal{K}_{\sigma } u _e^{\sigma }}, \label{EQ3.6}
\end{equation}
and
\begin{equation}
\mathcal{K}^{\mu }=\dot{x}_{\mu }. \label{EQQ3.6}
\end{equation}
In which,  $\mathcal{K}^{\mu }$  is the four-velocity of the photon, and $u_0^{\mu }=(1,0,0,0)$ is the four-velocity of the static observer. Moreover, the four-velocity of the infalling accretion correspond to $u_e^{\mu }$, which is
\begin{equation}
u_e^t={A (r)}^{-1},\quad u_e^r=-\sqrt{\frac{1-A (r)}{A (r) B ( r)}}, \quad u_e^{\theta }=u_e^{\varphi }=0.   \label{EQ3.7}
\end{equation}
Here, $\mathcal{K}_t$ is a constant of motion. For the photon,  $\mathcal{K}_r$  can  be obtained  by the equation $\mathcal{K}_{\mu } \mathcal{K}^{\mu }=0$, and consequently
\begin{equation}
\mathcal{K}_t=\frac{1}{b}, \quad \frac{\mathcal{K}_r}{\mathcal{K}_t}=\pm \sqrt{B (r) \left(\frac{1}{A (r)}-\frac{b^2}{r^2}\right)}. \label{EQ3.8}
\end{equation}
In which, the sign $+$ or $-$ corresponds to  the photon approaches to or goes away from the black hole. Hence, the redshift factor of the infalling accretion is
\begin{equation}
g_i=\left(u_e^t+\left(\frac{\mathcal{K}_r }{\mathcal{K}_e}\right)u_e^r \right) ^{-1}. \label{EQ3.9}
\end{equation}
On the other hand, the form of the proper distance should be rewritten as
\begin{equation}
d l_{prop}=\mathcal{K}_\mu u_e ^\mu d\lambda=\frac{\mathcal{K}_t}{g_i |\mathcal{K}_r|} d r. \label{EQ3.10}
\end{equation}
Considering  the specific emissivity has the same form as in the equation (\ref{EQ3.3}), the observed  flux for the distance observer  in the case of infalling spherical accretion model is
\begin{equation}
\mathcal{I}_{{obs}}({\nu^i_{o}})\propto \int _{\gamma _i}\frac{g_i^3  \mathcal{K}_t dr}{r^2 |\mathcal{K}_r|}.\label{EQ3.11}
\end{equation}
Based on the above equation, we also can  investigate the shadow  and observation luminosity  of the  Bardeen black hole  surrounded by  the infalling spherical accretion. Similarly, the observation luminosity distribution under different monopole charge $g$ is given, which is shown in Figure 4.
\begin{figure}[tbp]
\centering % \begin{center}/\end{center} takes some additional vertical space
\includegraphics[width=0.50\textwidth]{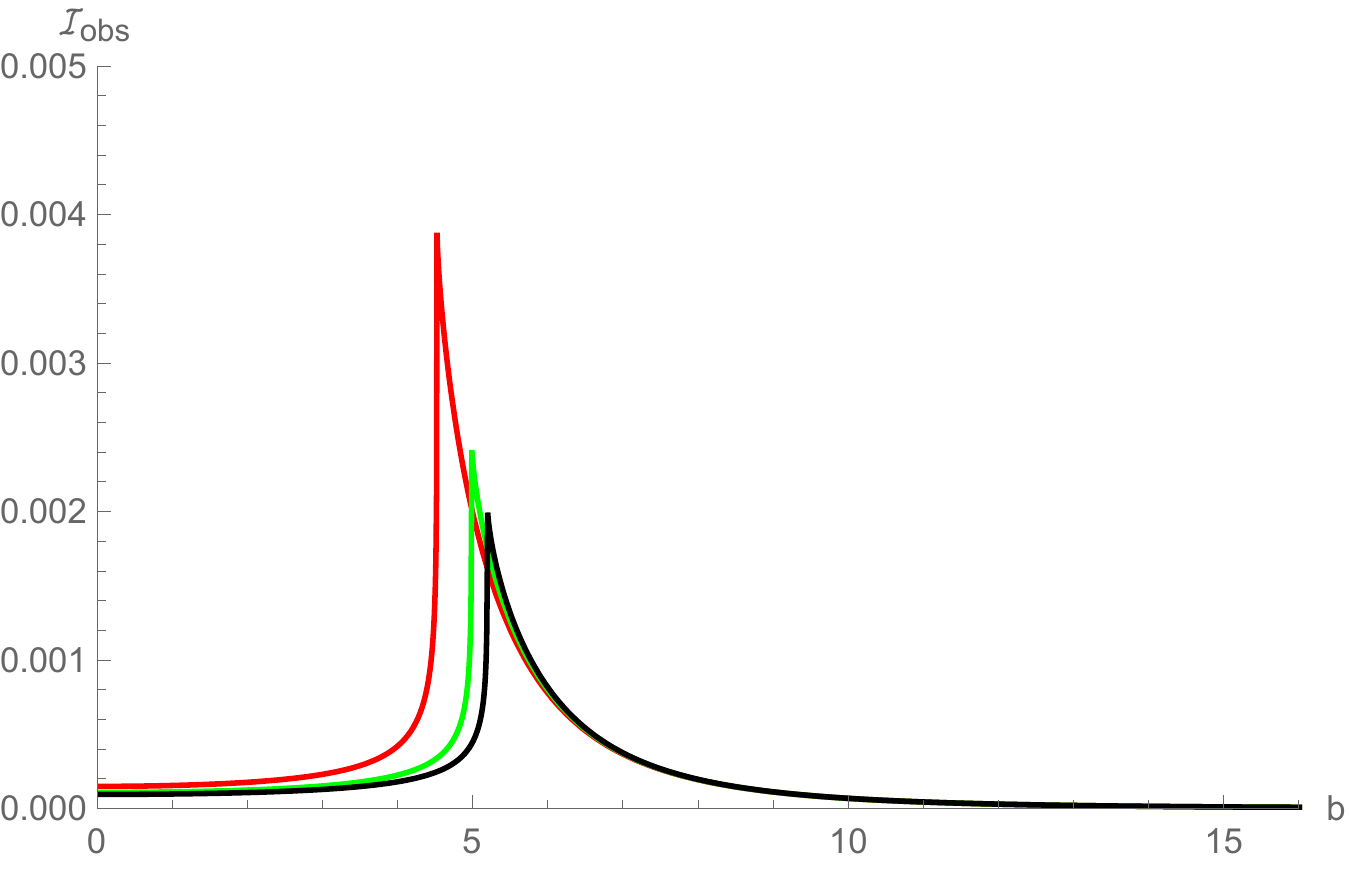}
% "\includegraphics" is very powerful; the graphicx package is already loaded
\caption{\label{fig4}  The observed specific intensity  of the black hole is surrounded by  the infalling spherical accretion, in which the black line, green line and red line correspond to $g=0$, $g=0.475$ and $g=0.75$, respectively. Here, we set M = 1. }
\end{figure}

One can find that the maximum value of observed intensity still at the position of  photon sphere ($b\sim b_p$). In addition, the observed intensity increases with the increase of $b$ in the region of $b < b_p$, and  decreases with the increase of $b$ in the region $b > b_p$.  However, the peak value of the observed  intensity is much smaller than that of the static model,  but the difference of peak value is more obvious under the different values of monopole charge $g$. In other words, the black hole shadow with the infalling spherical accretion is darker than that of the static case. The two-dimensional image of the observed intensity is also shown in Figure 5.

\begin{figure}[h]
\centering % \begin{center}/\end{center} takes some additional vertical space
\includegraphics[width=.325\textwidth]{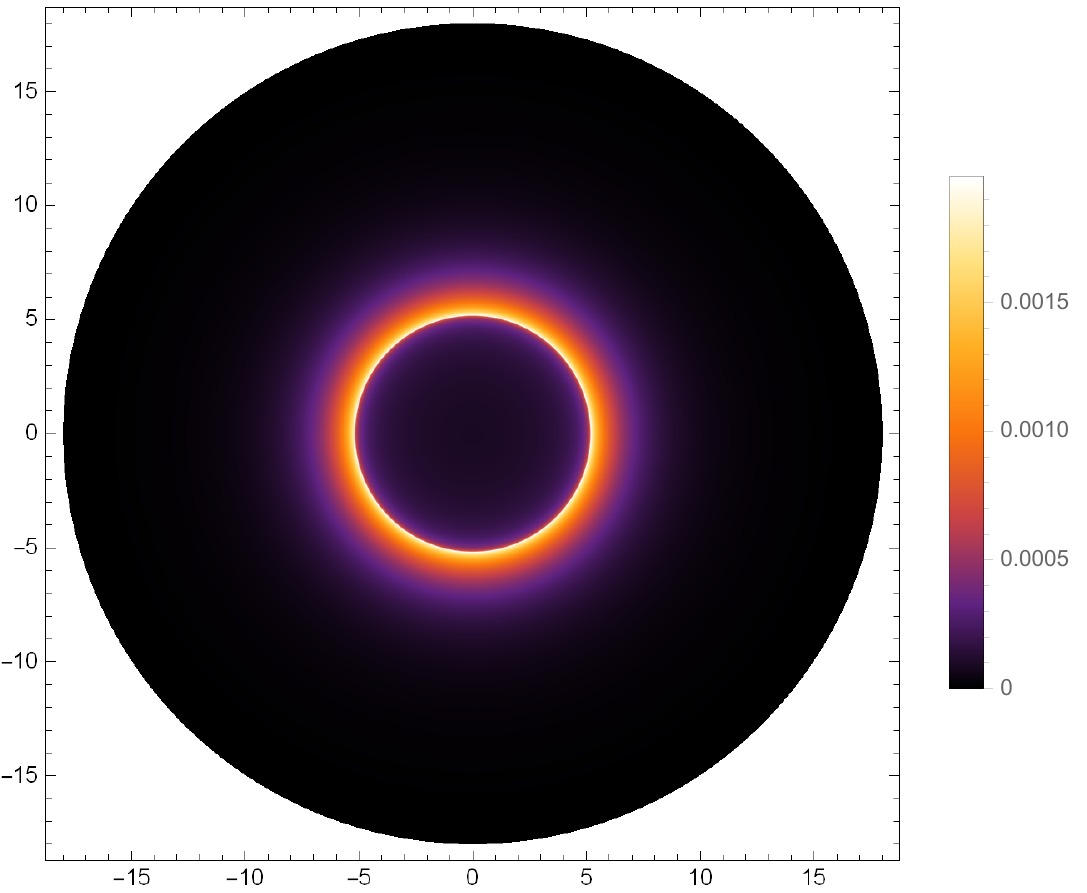}
\hfill
\includegraphics[width=.325\textwidth]{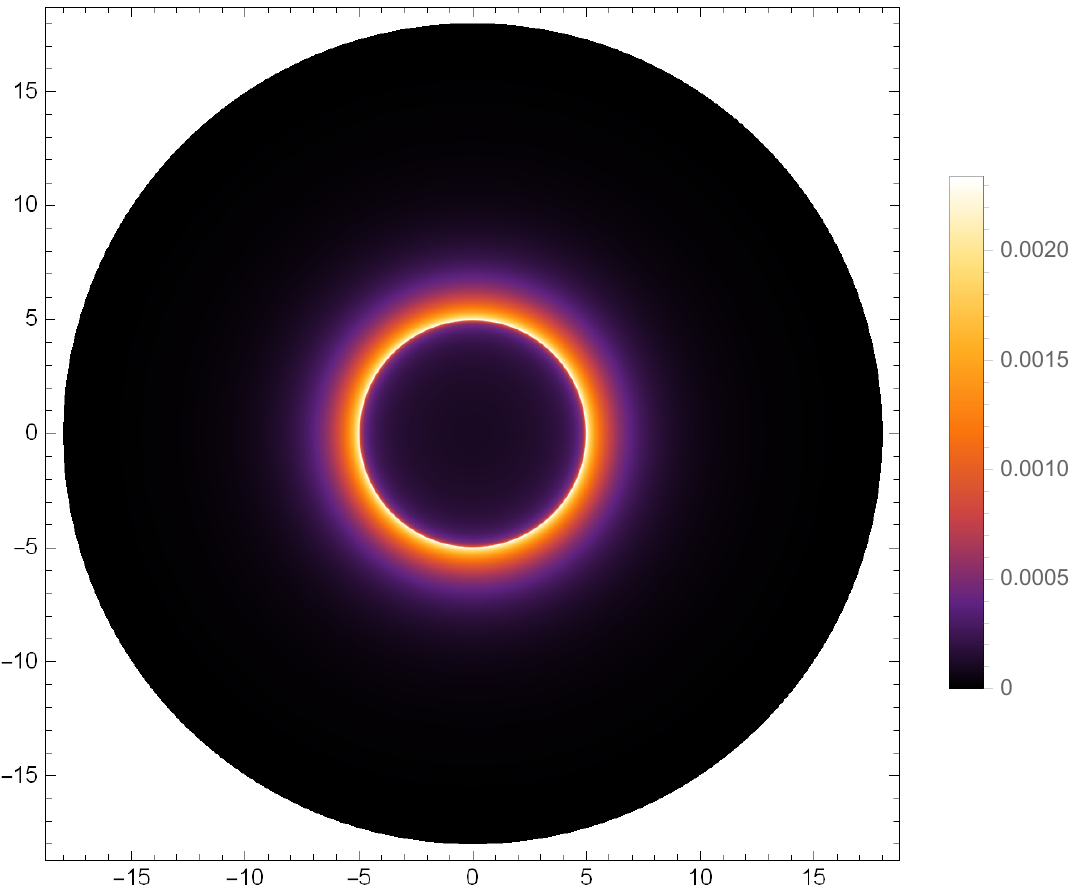}
\hfill
\includegraphics[width=.325\textwidth]{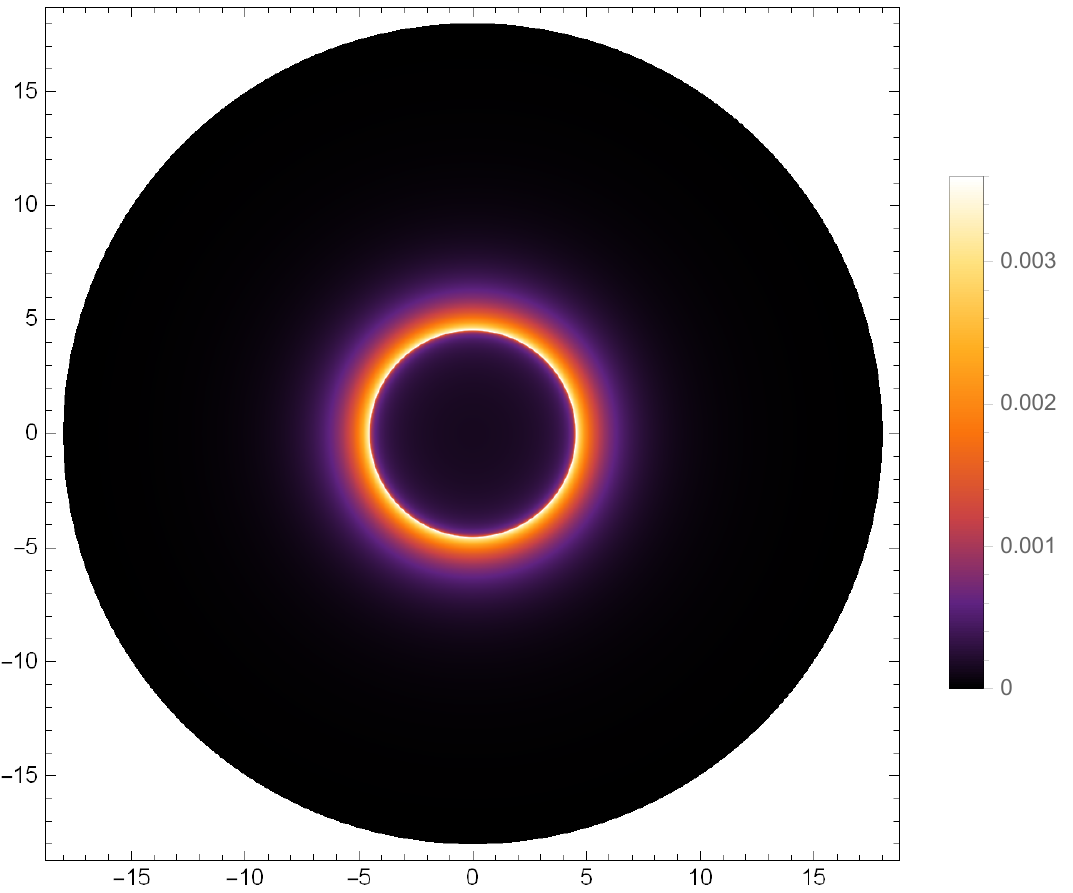}
\caption{\label{fig5}  Image of the black hole shadow with the infalling spherical accretion for the different values of monopole charge $g$, in which $g=0$ (left panel), $g=0.475$ (mid panel) and $g=0.75$ (right panel).  }
\end{figure}

Indeed, the central region inside the photon sphere in Figure 5 is  evidently  darker than the corresponding region in Figure 3, and the intensity of the bright ring is far less than that of static accretion.  The obvious difference between these two  models is due to the Doppler effect, as near the event horizon of the black hole, this effect is more obvious. In addition, it is can be  find that the radii of  shadow and the position of  photon sphere remain unchanged in the different  accretion progress. This suggests that  the  shadow is an inherent property of spacetime, the behavior of accretion flow around the black hole only affect the observation intensity.

\section{The shadow and rings with thin disk accretion flow}

\subsection{The behavior of the trajectories of light rays}
In this section, we are going to  study the  shadow when a black hole  surrounded by the thin disk accretion in  the Bardeen black hole, where the optically and geometrically thin disks is placed  on the equatorial plane of  black hole. Due to the trajectories of light ray  near the black hole are the important basis for studying the shadow and  observation feature of  black hole. In \cite{Gralla:2019xty},  Gralla et al. distinguish the trajectories of light rays near the black hole, and they put forward the total number of orbits which is defined as
\begin{equation}
n (b)=\frac{\varphi }{2 \pi },\label{EQ4.1}
\end{equation}
which is a function of impact parameter $b$.  Using equation (\ref{EQ4.1}),  the trajectories of the light ray  is divided into three types, namely, the direct, lensing and photon ring ones. That is, there are not only the photon rings but also lensing rings outside of the  black hole shadow. In the case of $n(b) < 3/4$, the light ray  will intersect the equatorial plane only once, corresponding to  the direct emissions. In the case of $3/4 < n(b) < 5/4$, the light ray crossing the equatorial plane at least twice, corresponding to  the lensing rings. In the case of $n(b) > 5/4$, the light ray  will intersect the equatorial plane at least three times, and it corresponds to the photon rings.
For the different values of monopole charge $g$, the range of impact parameter corresponding to direct, lensing rings  and photon rings in the Baedeen spacetime is shown in Table 2. And, we also show the relationship between the total number of orbits $n(b)$ and impact parameter $b$ in Bardeen spacetime, which is shown in Figure 6.
\begin{center}

{\footnotesize{\bf Table 2.} The region of direct, lensing rings and photon rings  corresponding to  impact parameters $b$ under the different values of monopole charge $g$, where $M=1$.\vspace{1mm}
\begin{tabular}{| c | c | c | c | }
\hline
  g     & Direct                              & Lensing rings                        & Photon rings\\
\hline
0       &$b<5.0267$ or $b>6.1927$     & $5.0267<b<5.1893$ and $5.2305<b<6.1927$   &$5.1893< b < 5.2305$  \\

\hline
0.1	    &$b<5.0049$ or  $b >6.1623$     & $ 5.0049<b<5.179$ and $5.2196<b<6.1623$  &$5.179< b < 5.2196$ \\

\hline
0.475    &$b<4.7564$ or $b>6.0446$     &$4.7564<b<4.9717$ and $5.0276<b<6.0446$   &$4.9717< b < 5.0276$  \\

\hline
0.75    &$b<4.0769$ or $b>5.8461$     &$4.0769<b<4.5142$ and $4.6561<b<5.8461$   &$4.5142< b <4.6561$  \\
\hline
\end{tabular}}
\end{center}

\begin{figure}[h]
\centering % \begin{center}/\end{center} takes some additional vertical space
\includegraphics[width=0.55\textwidth]{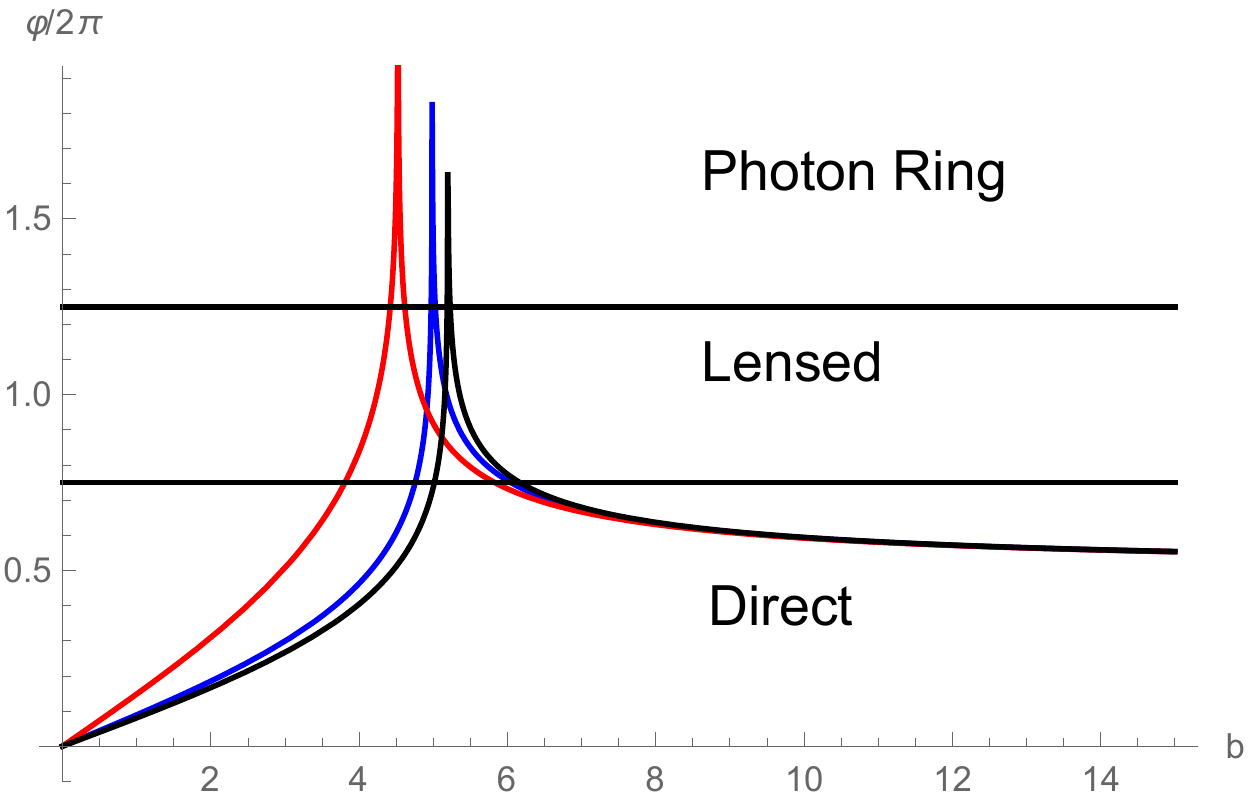}
% "\includegraphics" is very powerful; the graphicx package is already loaded
\caption{\label{fig6}  Behavior of photons in Bardeen black hole as a function of impact parameter $b$. The colors correspond to $g = 0$
(black), $g=0.475$ (blue), and $g=0.75$(red), respectively. }
\end{figure}

In Figure 6, it is found that the same value of $b$ may correspond to different regions under the different  values of monopole charge $g$,  which is corresponds to Table 2. In addition, the value of impact parameter $b$  corresponding to the  photon rings and lensing rings are smaller than  that of the Schwarzschild spacetime ($g=0$), but the existence of monopole charge $g$ leads to the increase of the ratio of the region occupied by the photon rings and lensing rings. When the impact  parameter $b$ increases to a large enough value, the trajectories  light ray is  direct case no matter what the value of $g$ is. That is, both the monopole charge $g$ and impact parameter $b$ will affect the observation characteristics of the Bardeen black hole. In order to more clearly distinguish the distribution of light ray trajectories near the black hole, we show the associated photon trajectories for Bardeen black hole in the polar coordinates ($b,\varphi$), which is shown in Figure 7.
\begin{figure}[h]
\centering % \begin{center}/\end{center} takes some additional vertical space
\subfigure[$g=0.1$]{\includegraphics[scale=0.35]{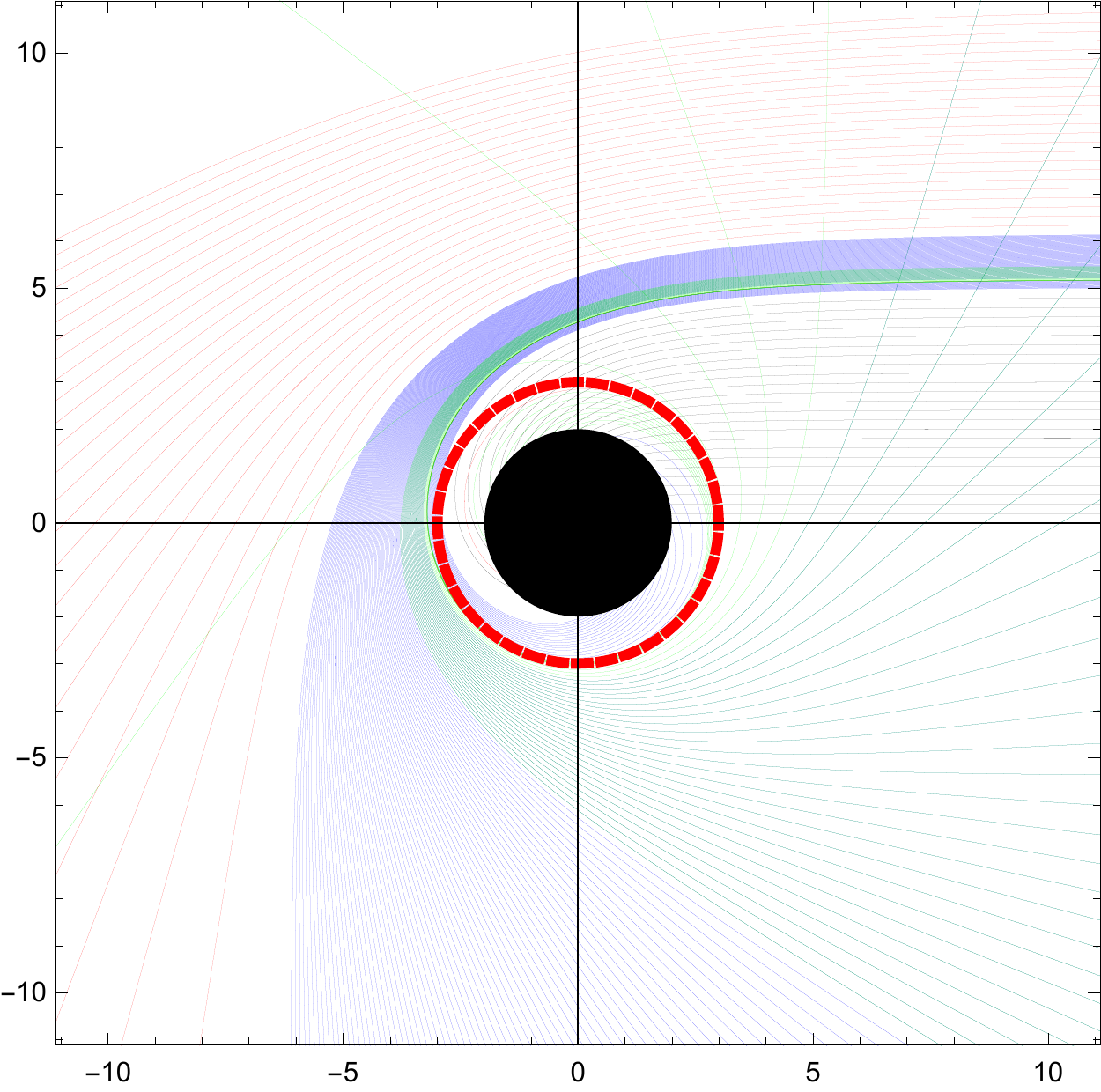}}
\subfigure[$g=0.475$]{\includegraphics[scale=0.35]{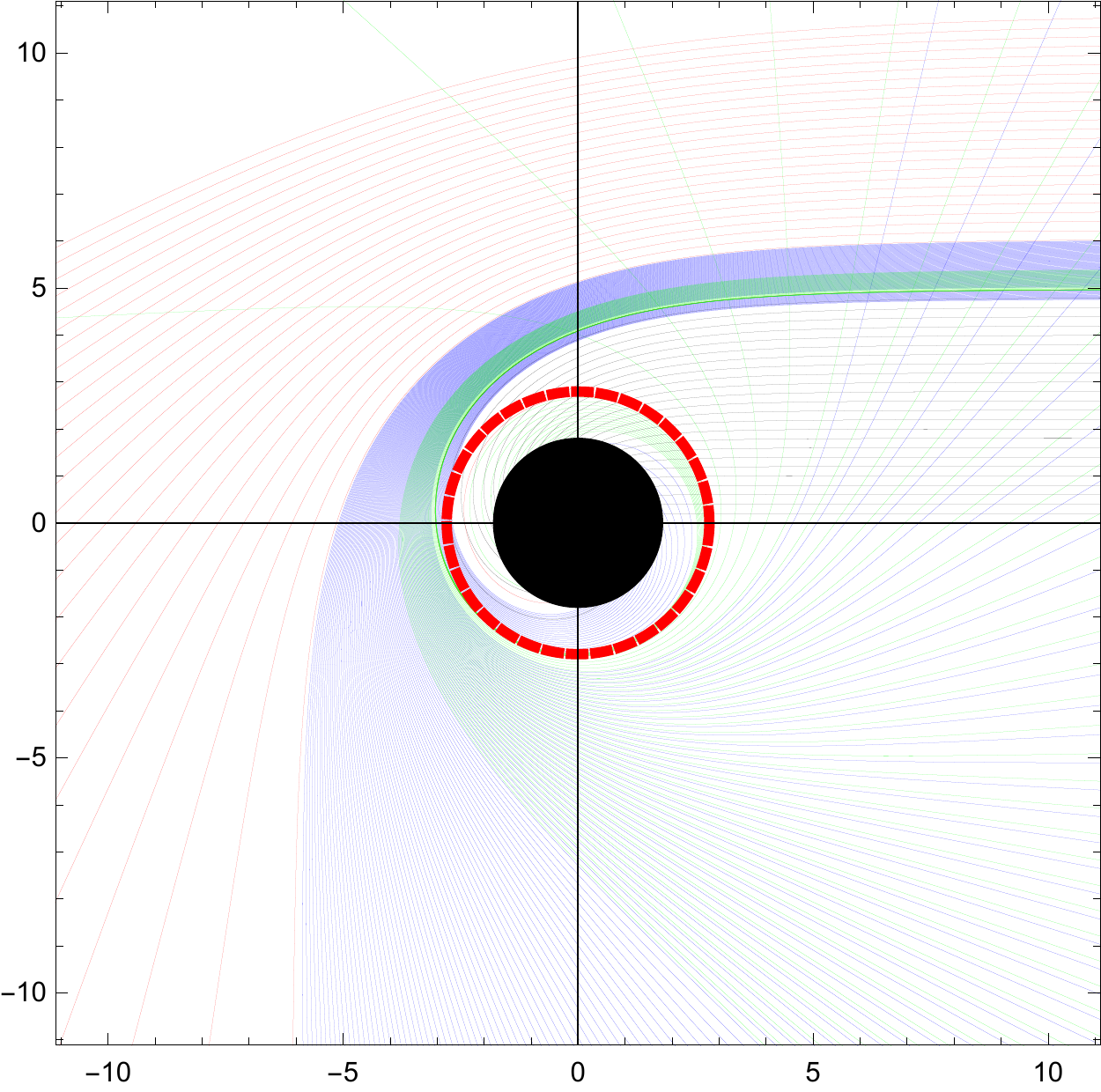}}
\subfigure[$g=0.75$]{\includegraphics[scale=0.35]{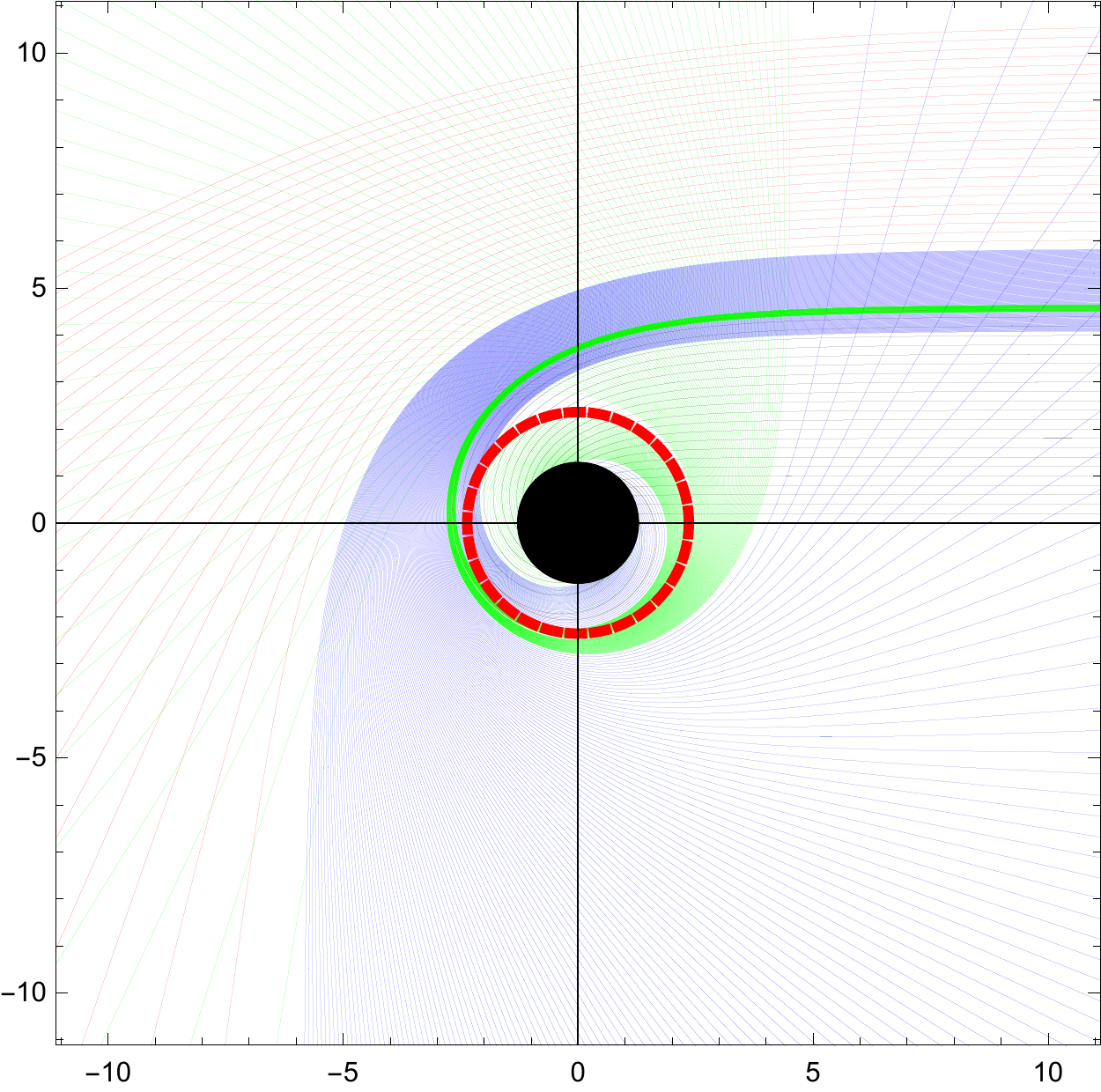}}
\caption{\label{fig8}  The selection of associated photon trajectories in the polar coordinates $(b, \varphi)$, in which the red lines, blue lines and green lines correspond to direct, lensing, and photon ring  bands, in which the spacing in impact parameter is 1/10, 1/100, and 1/1000 in the direct, lensing  and photon rings,  respectively. And, the black hoke are shown as the black disks,  the dashed red line represent the photon ring. }
\end{figure}

\subsection{Transfer functions}
We consider that a distance static observer is located at the north pole, and the  thin disk accretion at the equatorial plane of the black hole. Moreover, the lights emitted from the thin disk accretion is  isotropically for the static observer. In view of this, the specific intensity and frequency of the emission are expressed as $I_{\nu }^{{em}} (r) $ and $\nu_{em}$,  so as the observed specific intensity and frequency are defined as $I_{\nu'}^{{obs}} (r)$ and  $\nu_{obs}$. By considering ${I_{\nu }^{{em}}}/{\nu _e^3}$ is conserved along a ray from the Liouville theorem, that is, one can obtain a correlation between the observed specific intensity and the emissivity specific intensity \cite{Zeng:2020vsj}, that is
\begin{equation}
I_{\nu ' }^{{obs}}= (A (r))^{3/2} I_{\nu }^{{em}}(r).\label{EQ4.2}
\end{equation}
Thus,  we can get the total observation  and  emitted intensity by integrating for the full frequency, which reads as
\begin{equation}
I_{{O}}=\int  I_{\nu ' }^{{obs}} (r) \, d\nu_{obs} ' =\int  (A (r))^2 I_{\nu }^{{em}} \, d\nu_{em} =(A (r))^2 I_{\nu }^{{em}} (r),
\end{equation}
and
\begin{equation}
I_{{E}}=\int I_{\nu }^{{em}} (r) d \nu_{em}.  \label{EQQ4.3}
\end{equation}
It is worth mentioning that the intensity of light emitted only from the accretion disk, and  the absorption or reflection of light and other factors are not considered. When the  trajectory of light ray  passes through the accretion disk on the equatorial disk, a certain luminosity will be obtained, and transmitted to the observer at infinity. As discussed earlier, the different values for the  number of orbits $n(b)$ defined the number of times that the ray path pass through the equatorial plane of the black hole.
In the case of  $3/4<n(b)<5/4$,  the light ray  will bend around the black hole and intersect with the back of the thin disk after the first intersection with the thin disk. In this way, the light ray  will pass through the thin disk twice. When $n(b)>5/4$,  the light passes through the back of the thin disk and then through the front of the thin disk again, due to the light ray is more curved around the black hole, i.e., the light will pass through the thin disk  at least three times. Hence, the sum of the intensity at each intersection is the total intensity that the observer can observe, which is
\begin{equation}
I_{{O}}(r)=\sum _n (A (r))^2 I_{{E}}|_{r= r_n(b)}. \label{EQ4.4}
\end{equation}
Here, $r_n(b)$ is the transfer function, which represents the radial position of the $n_{th}$ intersection with the thin emission disk, which means that the  relationship between radial coordinate $r$ and impact parameter $b$ can be expressed by the  transfer function. In addition, $dr/d b$ corresponds to the slope of the transfer function, which is called the  demagnification factor.  Actually, the slope of the transfer function reflects the demagnified scale of the transfer function\cite{Gralla:2019xty,Zeng:2020vsj}. In  Figure 8,  the graph of the transfer functions  is given under the different values of monopole charge $g$.  For the case of  $g = 0$(Schwarzschild spacetime), please refer\cite{Gralla:2019xty}.

\begin{figure}[h]
\centering % \begin{center}/\end{center} takes some additional vertical space
\subfigure[$g=0.1$]{\includegraphics[scale=0.4]{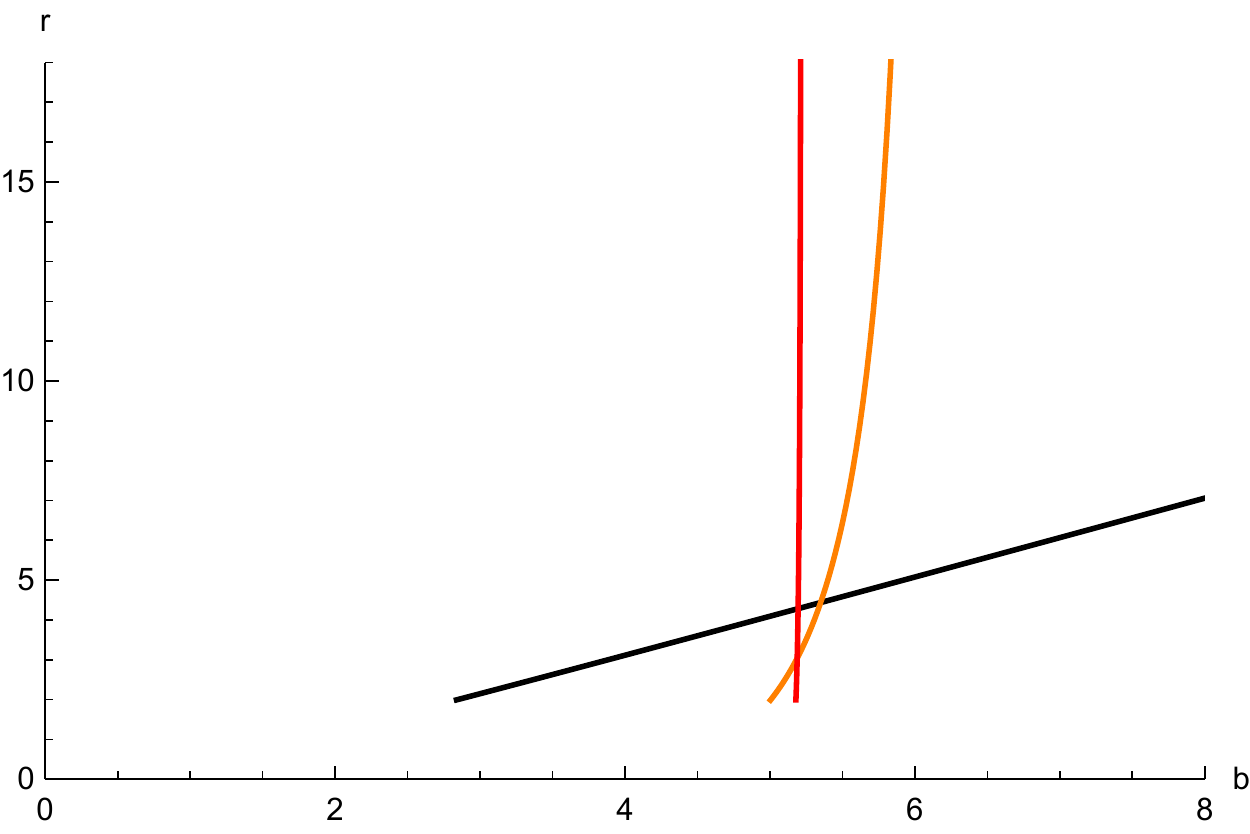}}
\subfigure[$g=0.475$]{\includegraphics[scale=0.425]{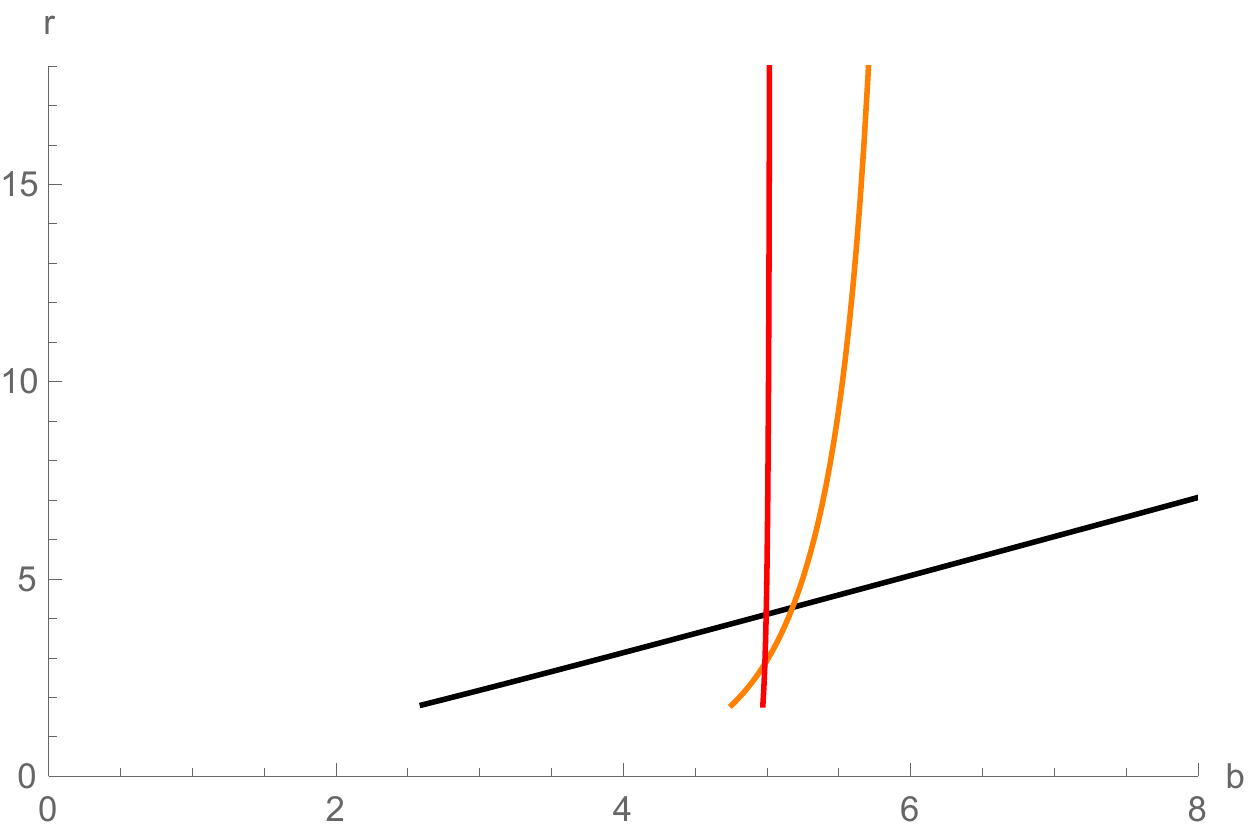}}
\subfigure[$g=0.75$]{\includegraphics[scale=0.425]{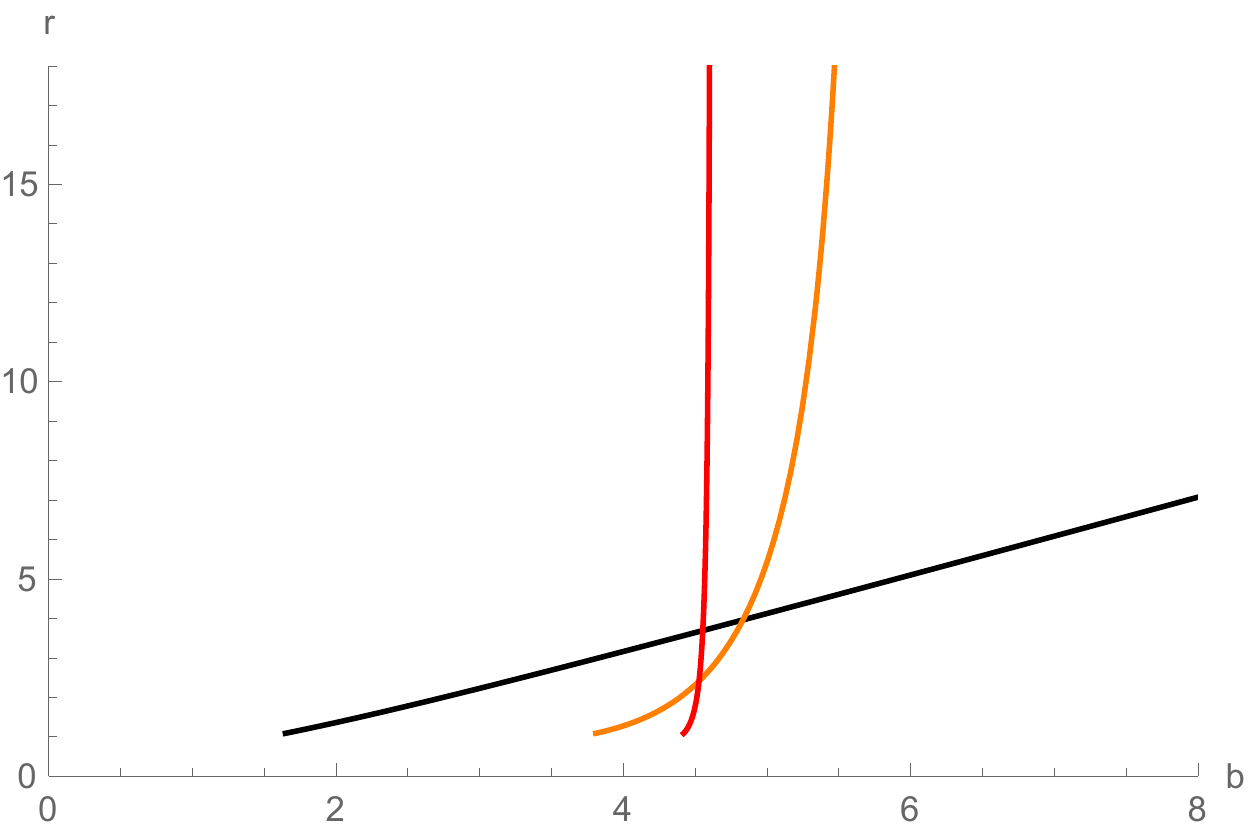}}
\caption{\label{fig8}  The relationship between the first three transfer functions $r_n(b)$ and $b$  in the Bardeen spacetime, in which the black, orange, and red lines correspond to the direct, lensing rings and
photon rings, respectively.  }
\end{figure}

The black line corresponds to the first transfer function($n=1$), and its slope ($\frac{d r}{db}|_1$) is a small fixed value. And, the first transfer function  corresponds to the direct image of the thin emission disk, which is the redshift of the source profile. The orange line corresponds to the second transfer function ($n=2$), on can find that the slope ($\frac{d r}{d b}|_2$) is a very large value. Therefore, the  observer will obtain  a highly demagnified image on the back of the thin disk, which  corresponds to the lensing rings. The red line represents the third transfer function ($n=3$),  whose slope ($\frac{d r}{d b}|_3$) is close to infinity($\frac{d r}{b d}|_3 \sim \infty$). Namely, the observer  will see an extremely demagnified image of the front side of the disk, and  the third transfer function corresponds to the photon ring. As a result, the observation luminosity provided by the photon rings and lensing rings only accounts for a small proportion of the total observed flux, the direct emission is the main measure of observed intensity. In addition, the image provided by the later transfer function is more demagnetized and negligible, so we only consider the first three transfer functions.

\subsection{The observational appearance of  Bardeen black hole surrounded by a thin disk accretion}

With the help of  preparatory work, we can further study the specific intensity of emission. For the emission form, we mainly consider three different models. In the Model I, the peak of emission intensity at the position of the innermost stable circular orbit ($r_{isco}$). There will be no emission inside the  innermost stable circular orbit, but exist a sharp downward trend outside the innermost stable circular track, which is
\begin{align}
    I_{E}(r_1) =\begin{cases}\left(\frac{1}{r-(r_{isco}-1)}\right)^2, &  r>r_{{isco}}  \\
    0, &r < r_{{isco}} \label{EQ4.5}
    \end{cases}
\end{align}
In the Model II, we  assume that the position of the emission peak located at the photon sphere ($r_p$), and the emission is a decay function of the power of third, that is
\begin{align}
    I_{E}(r_2) =\begin{cases}\left(\frac{1}{r-(r_{p}-1)}\right)^3, &  r> r_{p}   \\
    0, &r < r_{p} \label{Eq.17}
    \end{cases}
\end{align}
In the Model III,  there is  the decaying gradually of the emission  between the event horizon $r_h$ and  the innermost stable circular orbit $r_{isco}$, such as
\begin{align}
    I_{E}(r_3) =\begin{cases} \frac{1-\tan^{-1}(r-5)}{1-\tan^{-1}(r_h-5)}
    , &  r>r_h  \\
    0,  &r < r_h  \label{Eq.18}
    \end{cases}
\end{align}

Taking the monopole charge $g = 0.475$ as an example,  the results about the intensity of the emission and observation in Figure 9 are obtained. In Figure 9,  the left column is the change trend of emission profiles $I_E(r)$ as the increases of radius $r$, The middle column, which is  shown that the relationship between the observed intensity $I_O(r)$ and the impact parameter $b$. And,  the right column is the two-dimensional  density plots of observed intensities $I_O(r)$ in celestial coordinates. In addition, the first, second and third row correspond to the emitted model I, II and III, respectively. As can be readily seen from Figure 9,  there is a disparity for both the central black area  and observation intensity in the  different  emission models.
\begin{figure}[h]
\centering % \begin{center}/\end{center} takes some additional vertical space
\includegraphics[width=.35\textwidth]{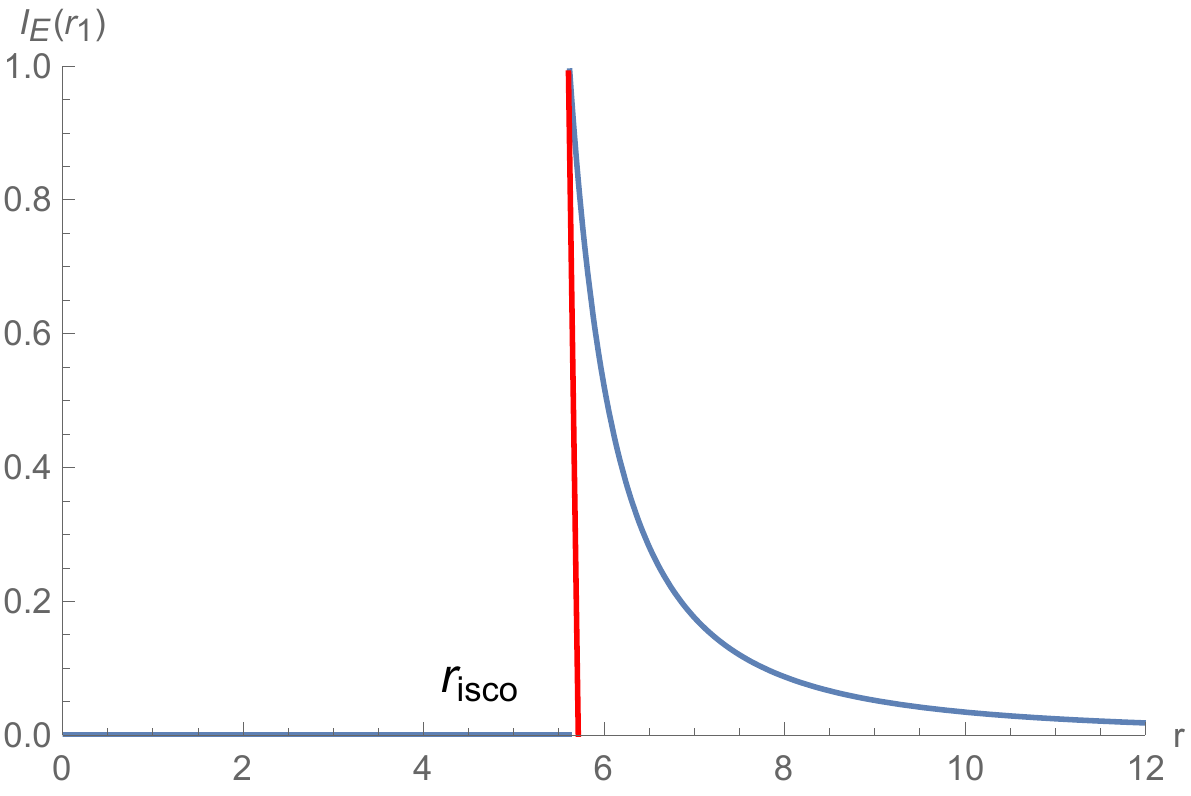}
\includegraphics[width=.35\textwidth]{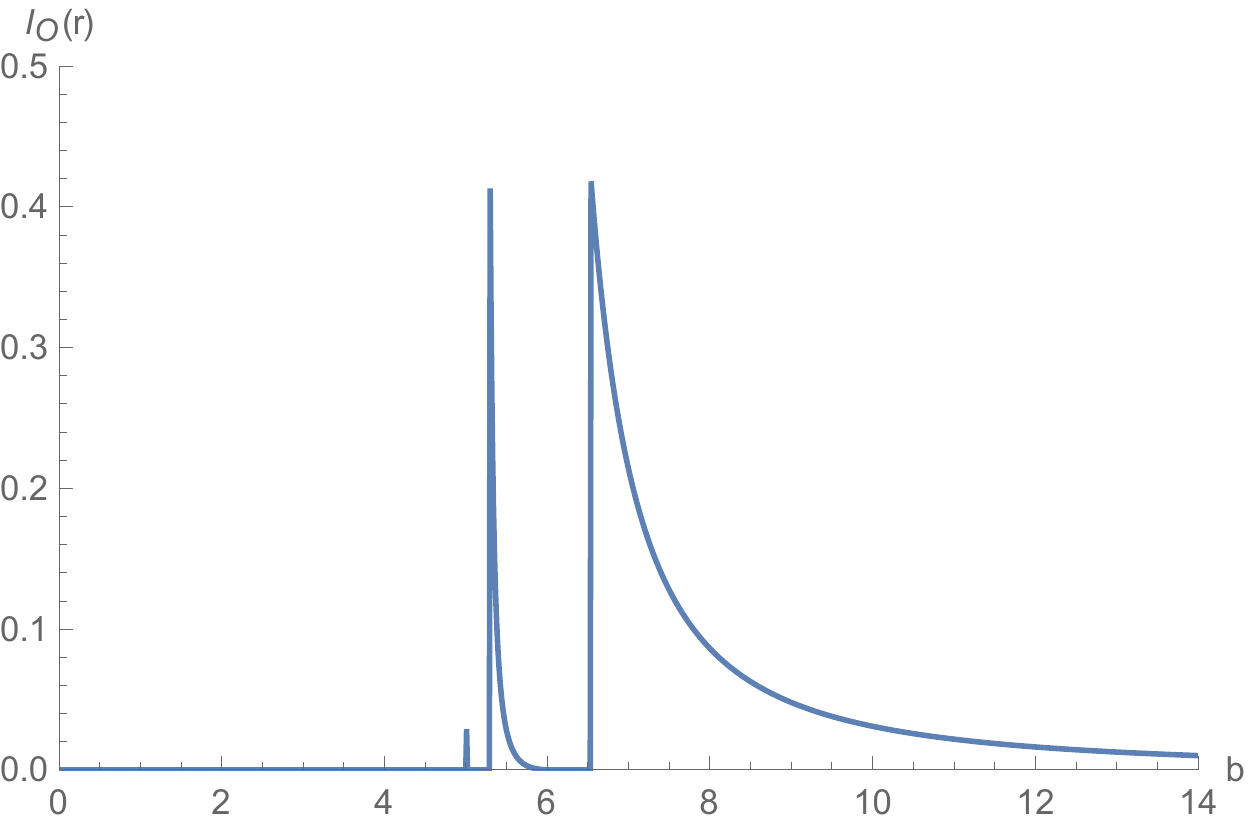}
\includegraphics[width=.28\textwidth]{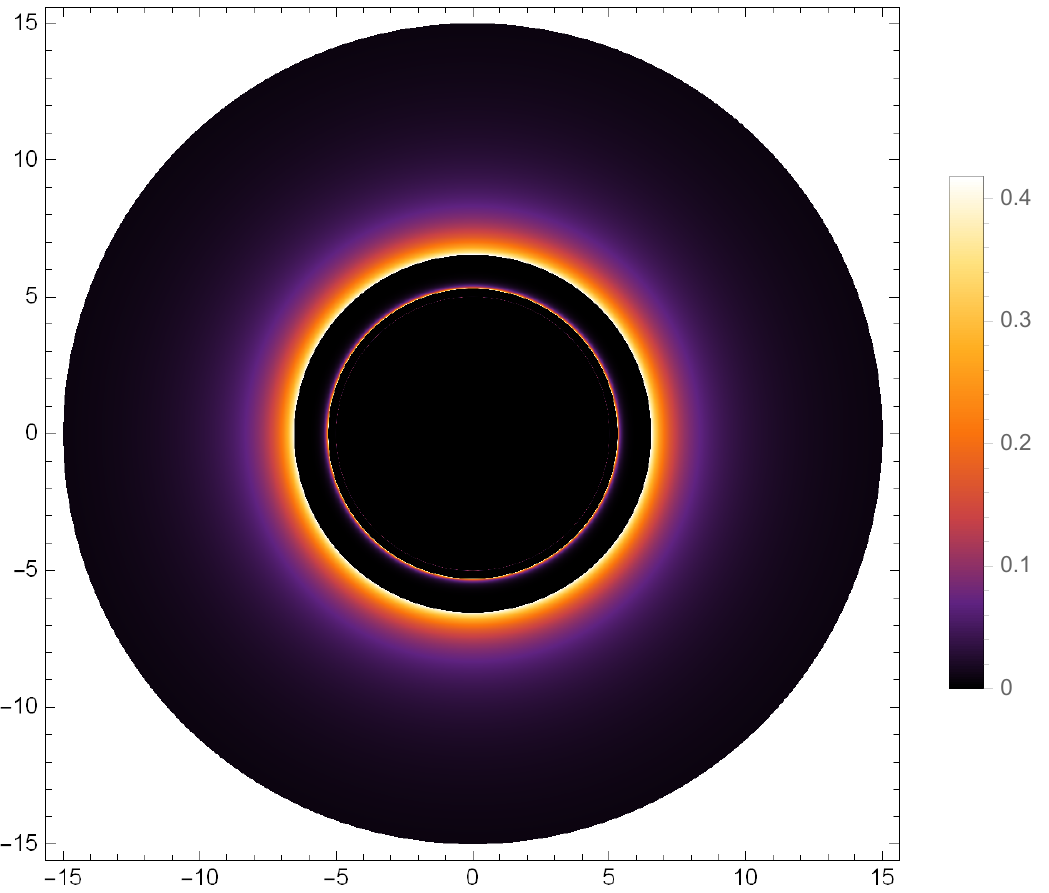}
\includegraphics[width=.35\textwidth]{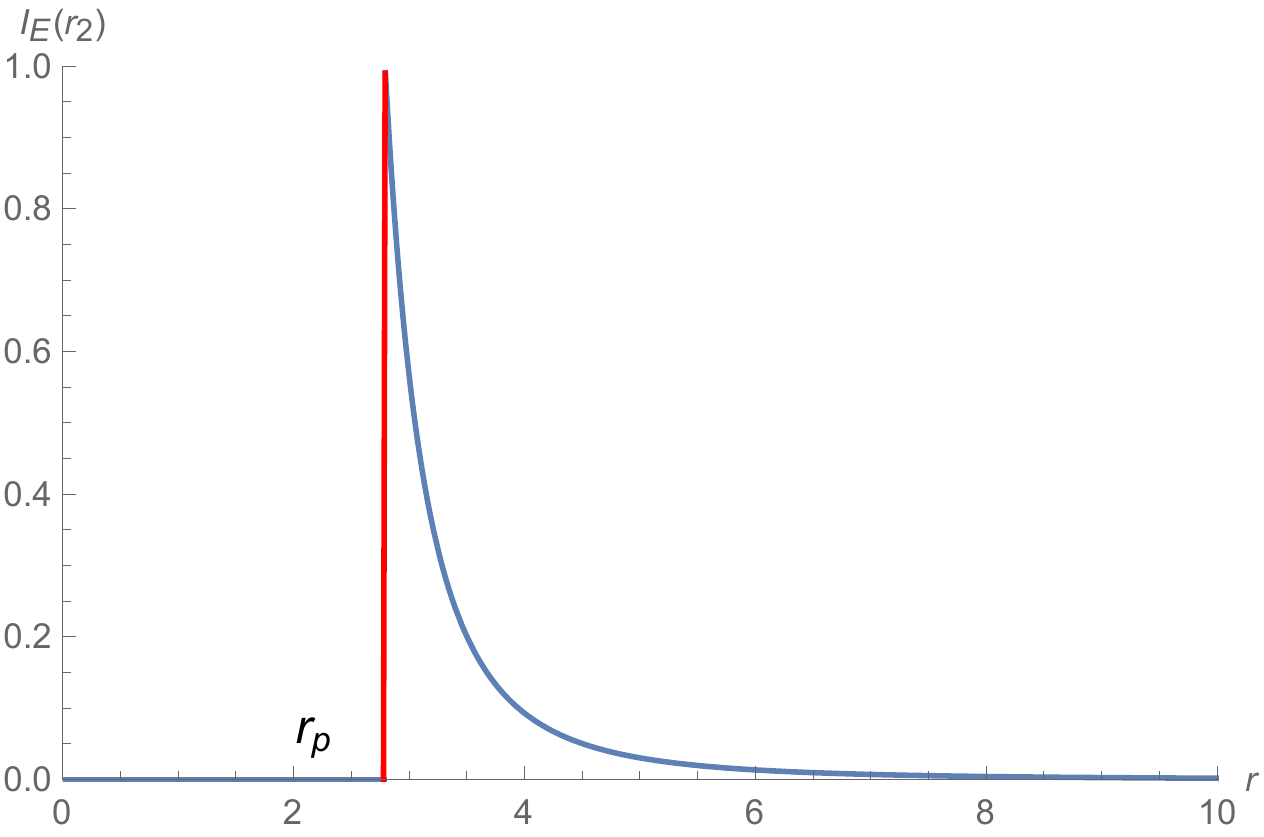}
\includegraphics[width=.35\textwidth]{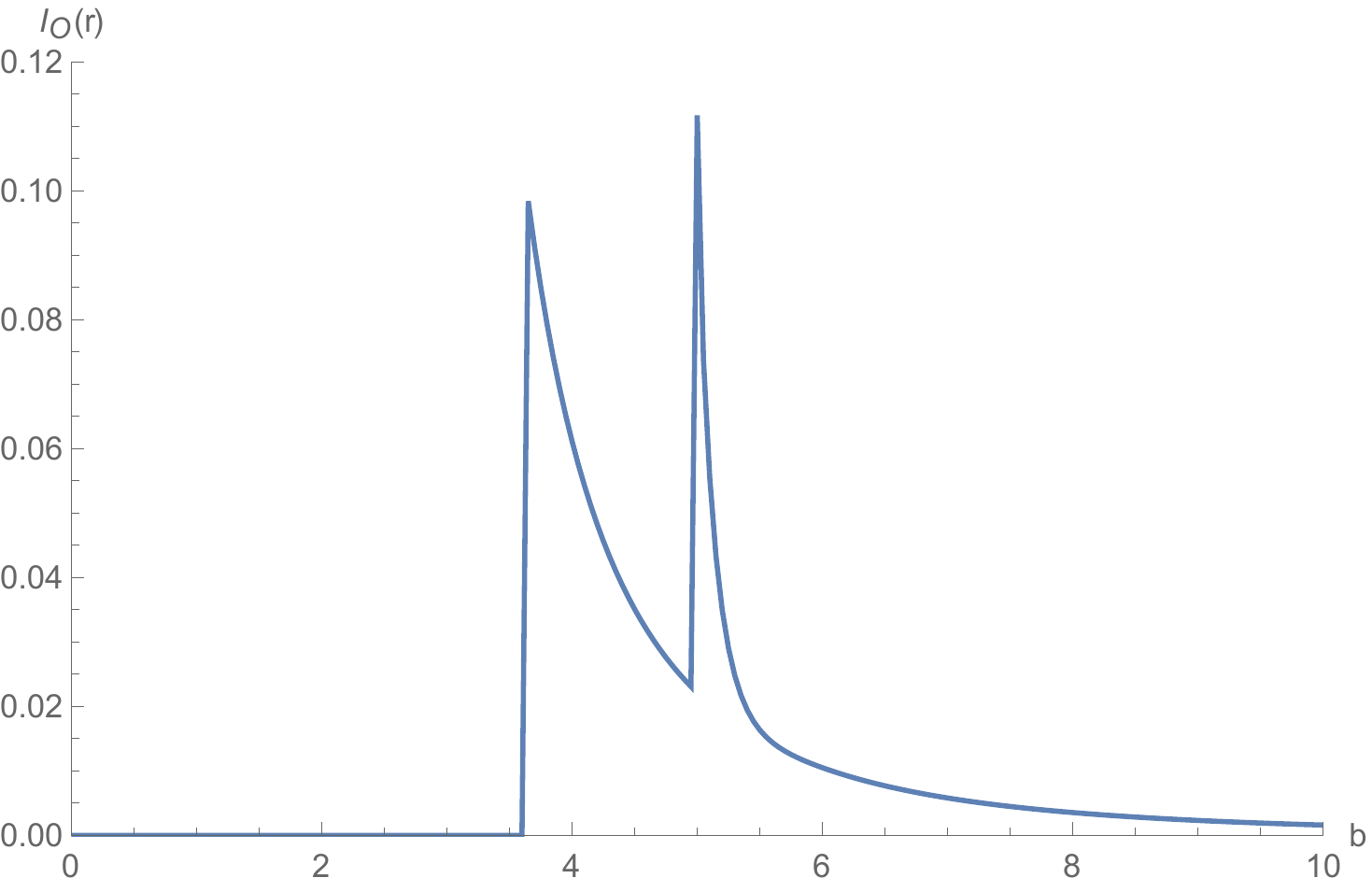}
\includegraphics[width=.28\textwidth]{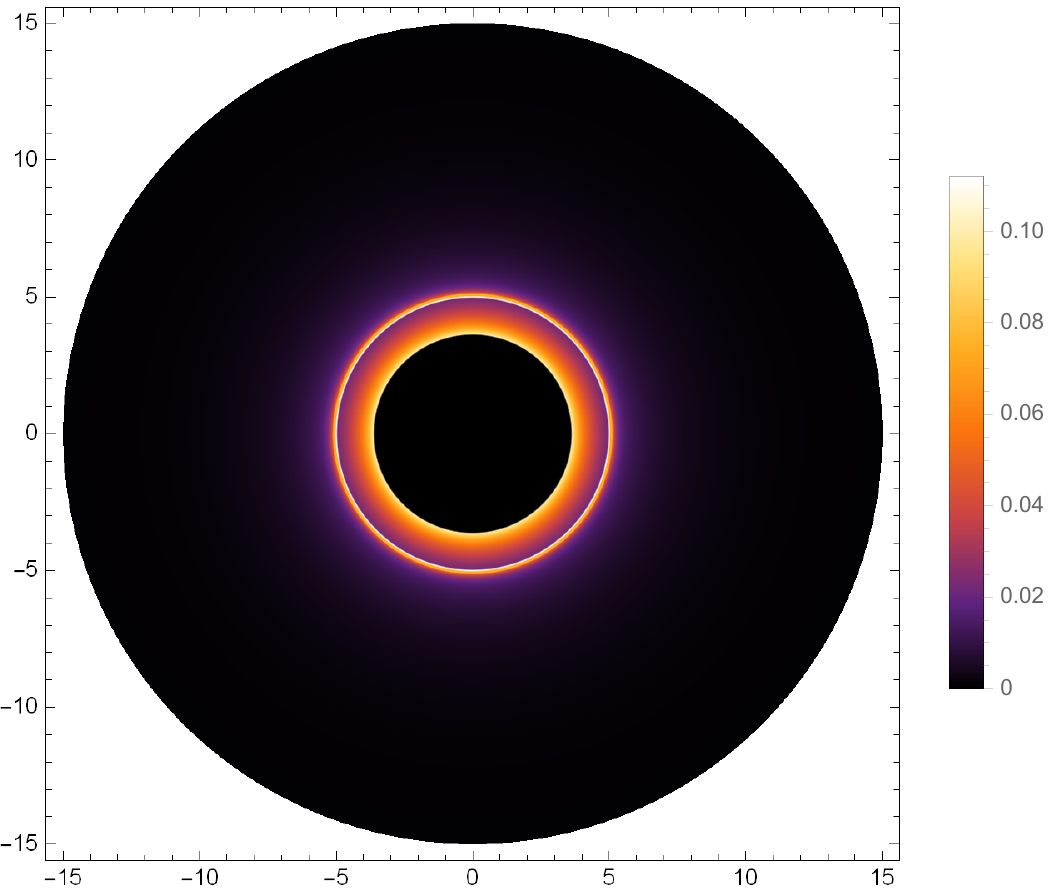}
\includegraphics[width=.35\textwidth]{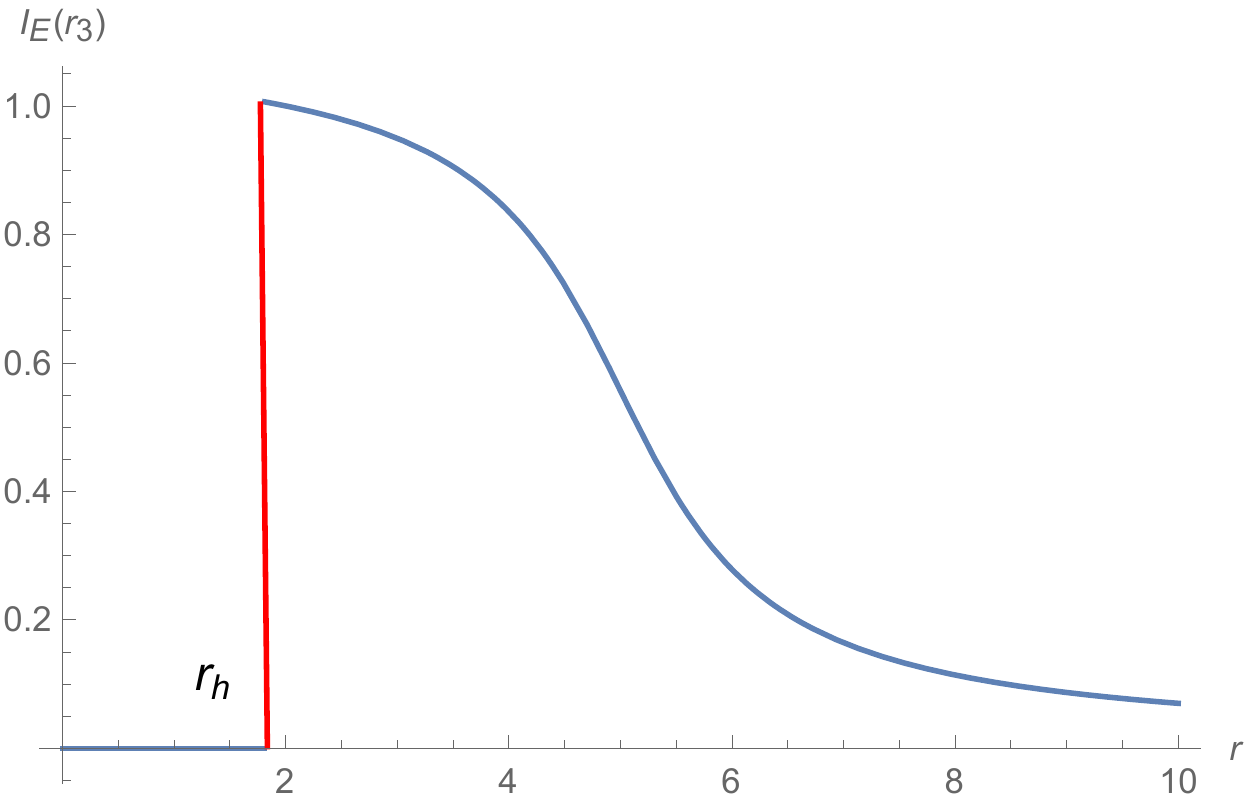}
\includegraphics[width=.35\textwidth]{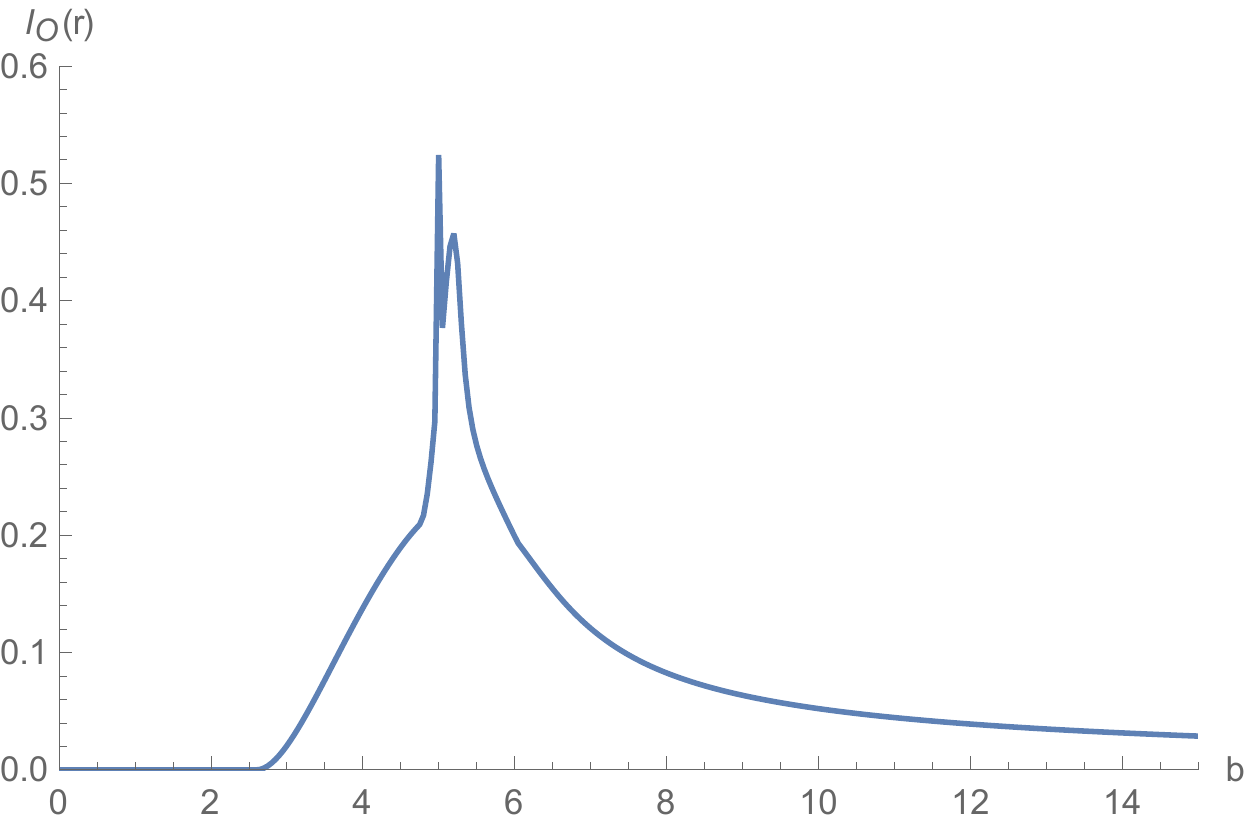}
\includegraphics[width=.28\textwidth]{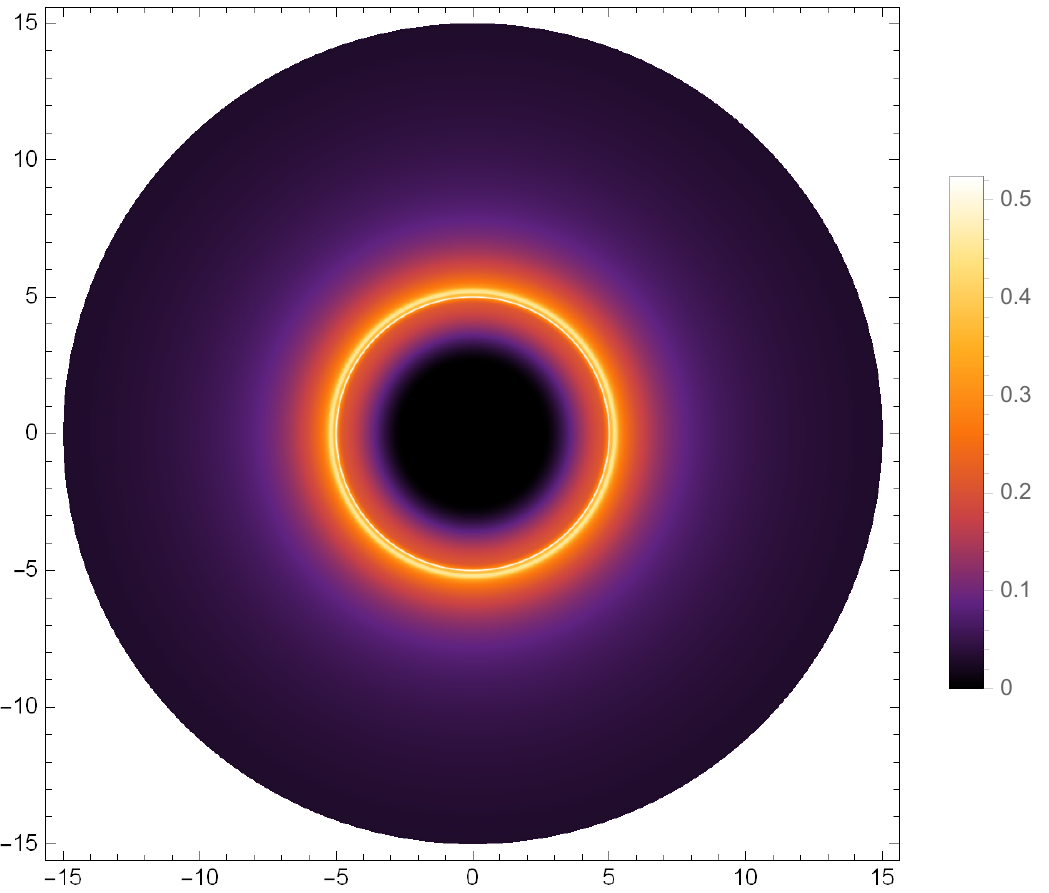}
% "\includegraphics" is very powerful; the graphicx package is already loaded
\caption{\label{fig10}  Observational appearance of the thin disk with different emission profiles  near the black hole, viewed from a face-on orientation. In which, the monopole charge $g=0.475$ and $M=1$.}
\end{figure}

In the Model I $I_{E}(r_1)$ where the emission starts with the position of  innermost stable circular orbit, although the photon rings can get more brightness from the thin disk, it is highly demagnetized.  Which can be see from the image of middle column, the observed intensity in the photon ring region is very small and  confined to a narrow region. Moreover, the two-dimensional density map also shows that there is an almost invisible thin line in the photon ring region. Therefore, the contribution of the photon ring to the total observed flux is negligible. For the lensing rings,  the observed intensity is larger than that of the photon rings, but the area occupied by the lensing rings is also very narrow, which leads to a bright line in the area of the lensing rings in the two-dimensional density map. As the result, the observation intensity obtained by the observer at infinity is mainly provided by the direct emission.

In the  Model II $I_{E}(r_2)$ where the emission starts at the   photon sphere, the emission regions of the lensing rings and the photon rings overlap, which resulting in a bright band outside the shadow has been shown in the two-dimensional density map. The contribution of the lensing rings to the total observed flux is higher than that of model I, while the contribution of the photon ring is still negligible. That is to say, the observation intensity obtained by the distant observer still directly determined by the direct emission.

In the  Model III $I_{E}(r_3)$ where the emission starts from the event horizon of black hole, the observation intensity increases with the increase of $b$ from the outer edge of  event horizon until it reaches to the peak near the lensing ring. The photon rings and  lensing ring completely coincide and cannot be distinguished.  In addition, the contribution of the lensing rings to the total observed flux is partially helpful and the photon ring also makes a negligible contribution, but the direct emission remains to dominate the total observed intensity.

Then, we also  plot the profiles of the emission and observed intensities when the monopole charge $g=0.75$, which is shown in Figure 10. By comparing Figure 9 with Figure 10, it can be found that the observation intensity,  range of the center dark area and  size of the light band outside  shadow are all  different obviously. When the value of monopole charge $g$  increases, the range of the center dark area and  size of the light band outside  shadow  decrease. Moreover, the observation intensity obtained by the distance observer  also decreases. Due to  the proportion of photon rings and lensing rings area increases with the larger monopole charge $g$, and the proportion of direct emission area decreases relatively, which naturally leads to the weakening of the observation intensity determined by direct emission. Therefore, the observed intensity of the light band outside  shadow is darker  than that of  the Schwarzschild spacetime, which could be an effective characteristic to distinguish the Bardeen black hole  from  Schwarzschild black hole.

\begin{figure}[h]
\centering % \begin{center}/\end{center} takes some additional vertical space
\includegraphics[width=.35\textwidth]{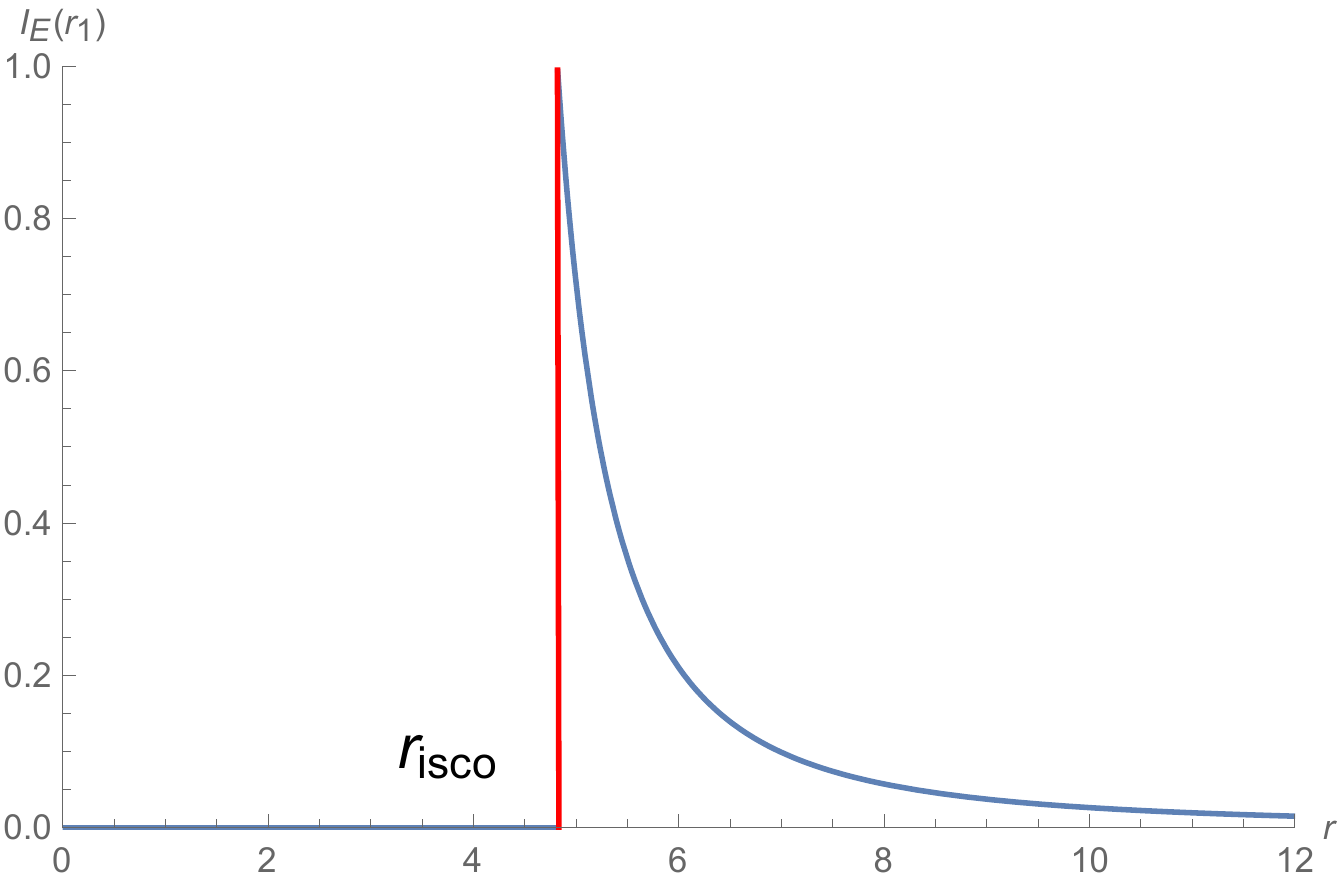}
\includegraphics[width=.35\textwidth]{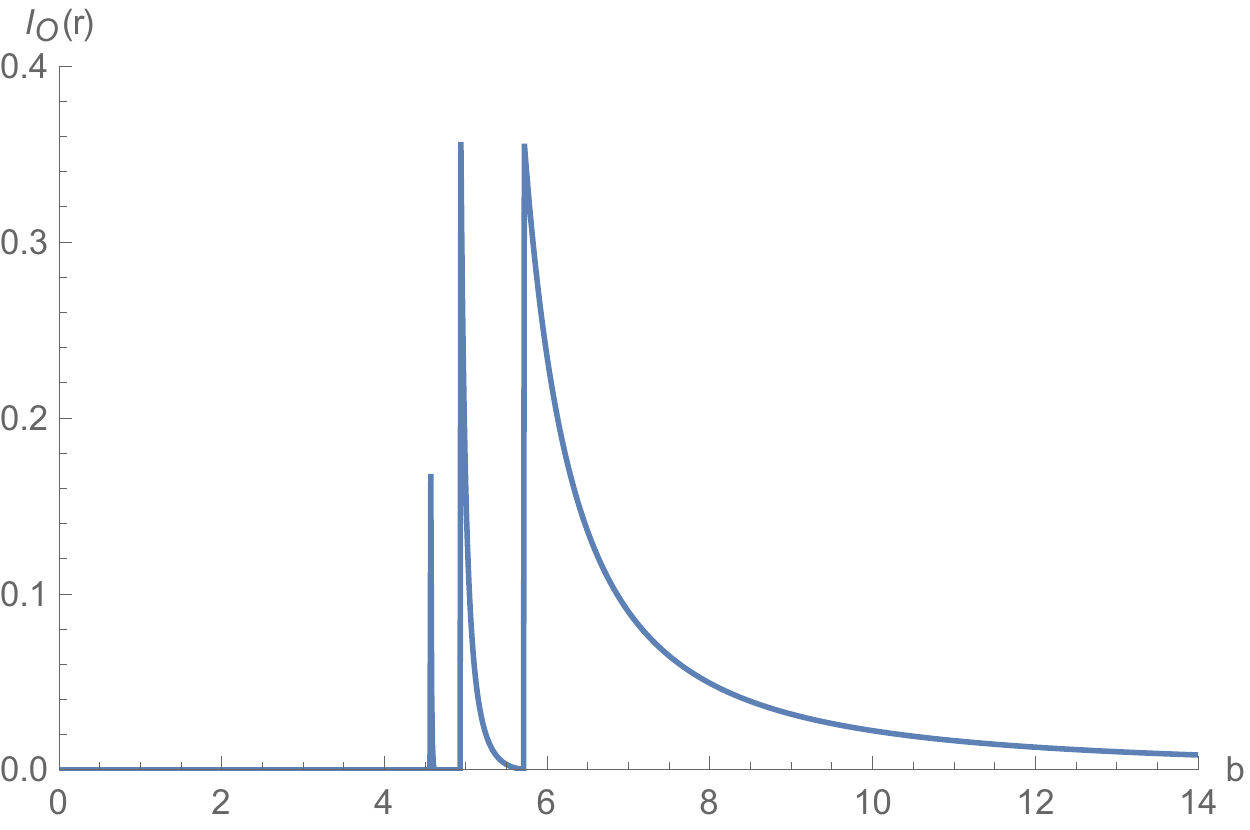}
\includegraphics[width=.28\textwidth]{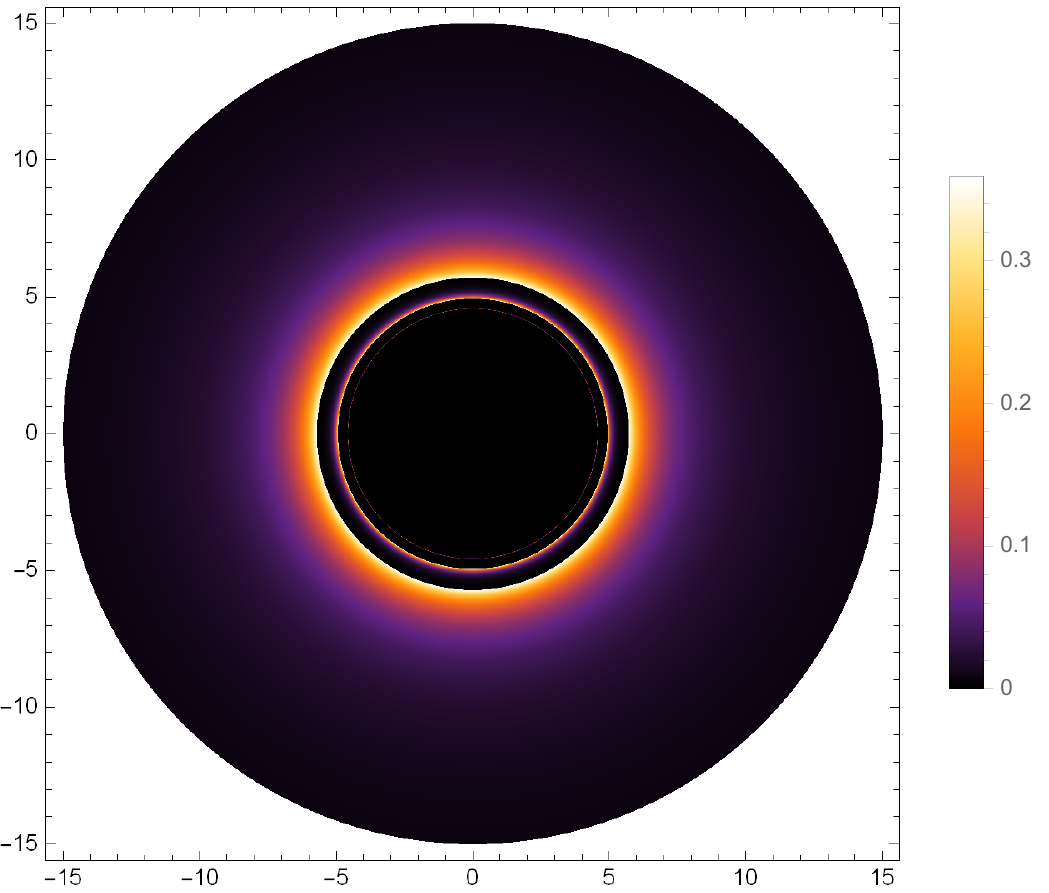}
\includegraphics[width=.35\textwidth]{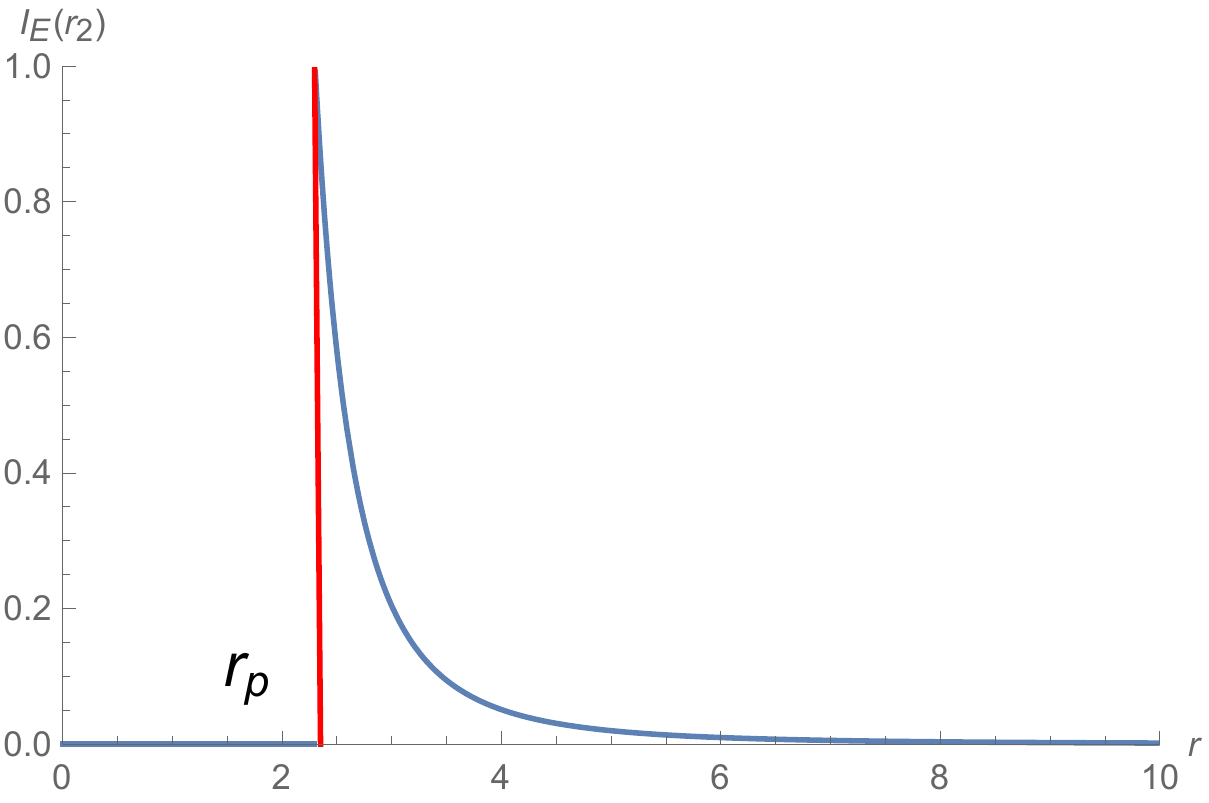}
\includegraphics[width=.35\textwidth]{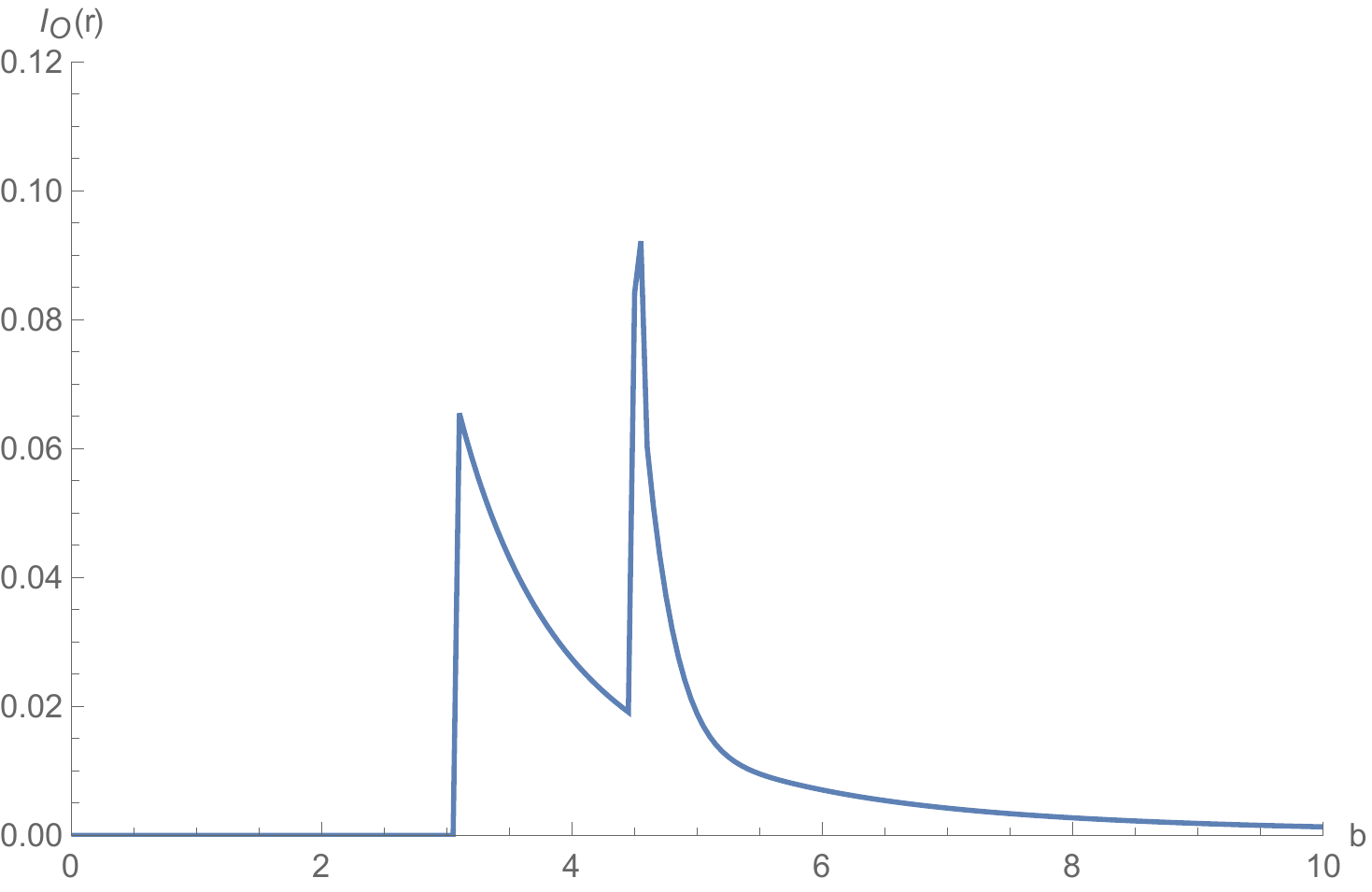}
\includegraphics[width=.28\textwidth]{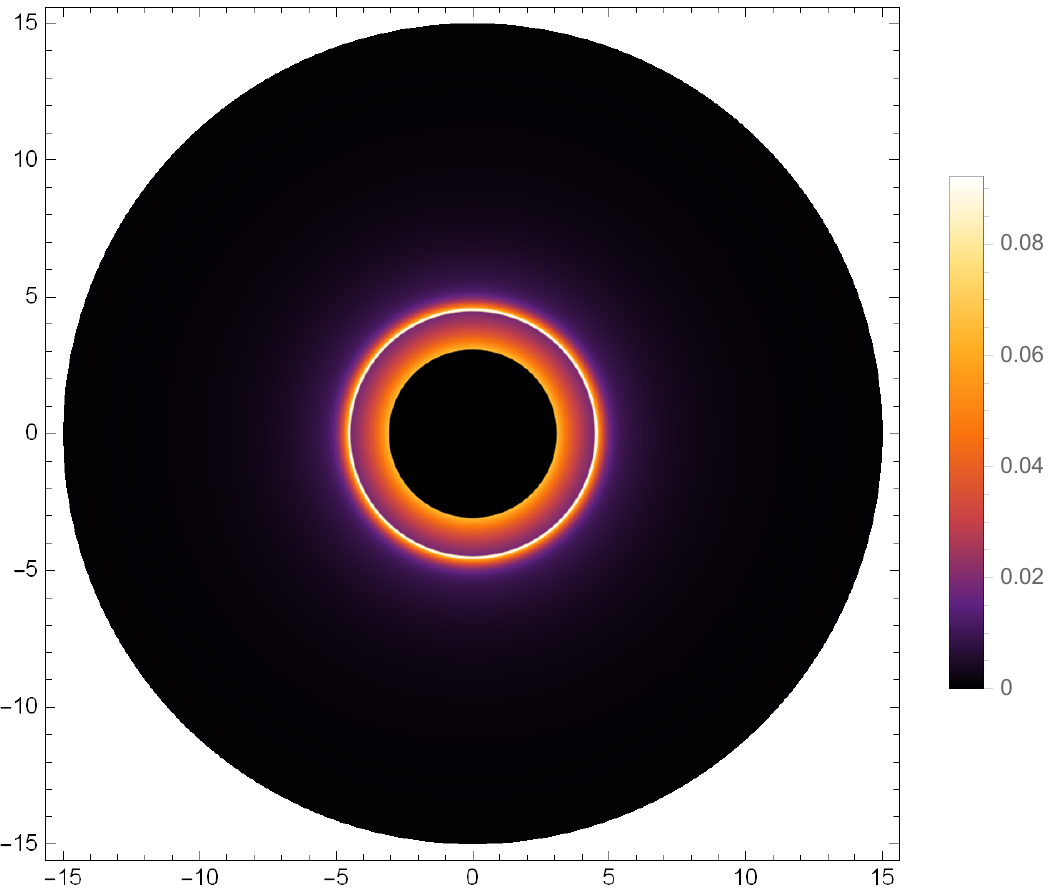}
\includegraphics[width=.35\textwidth]{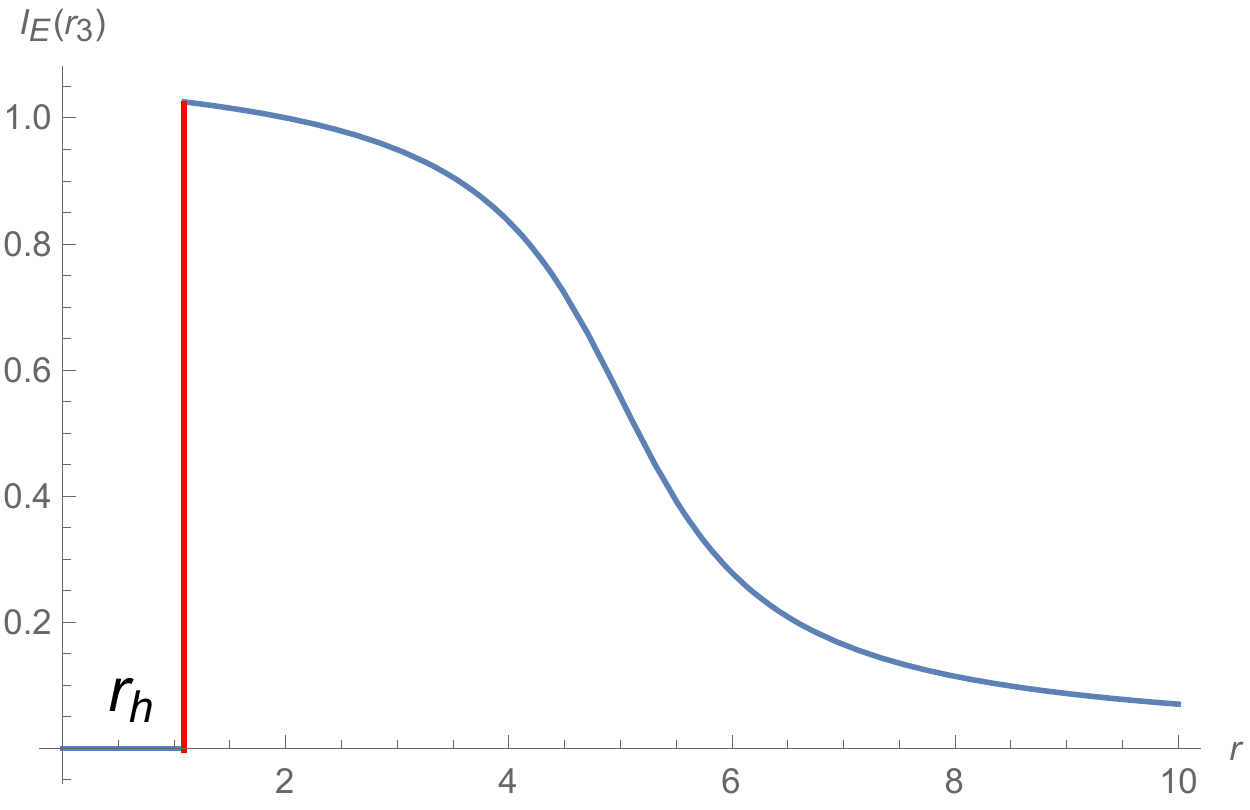}
\includegraphics[width=.35\textwidth]{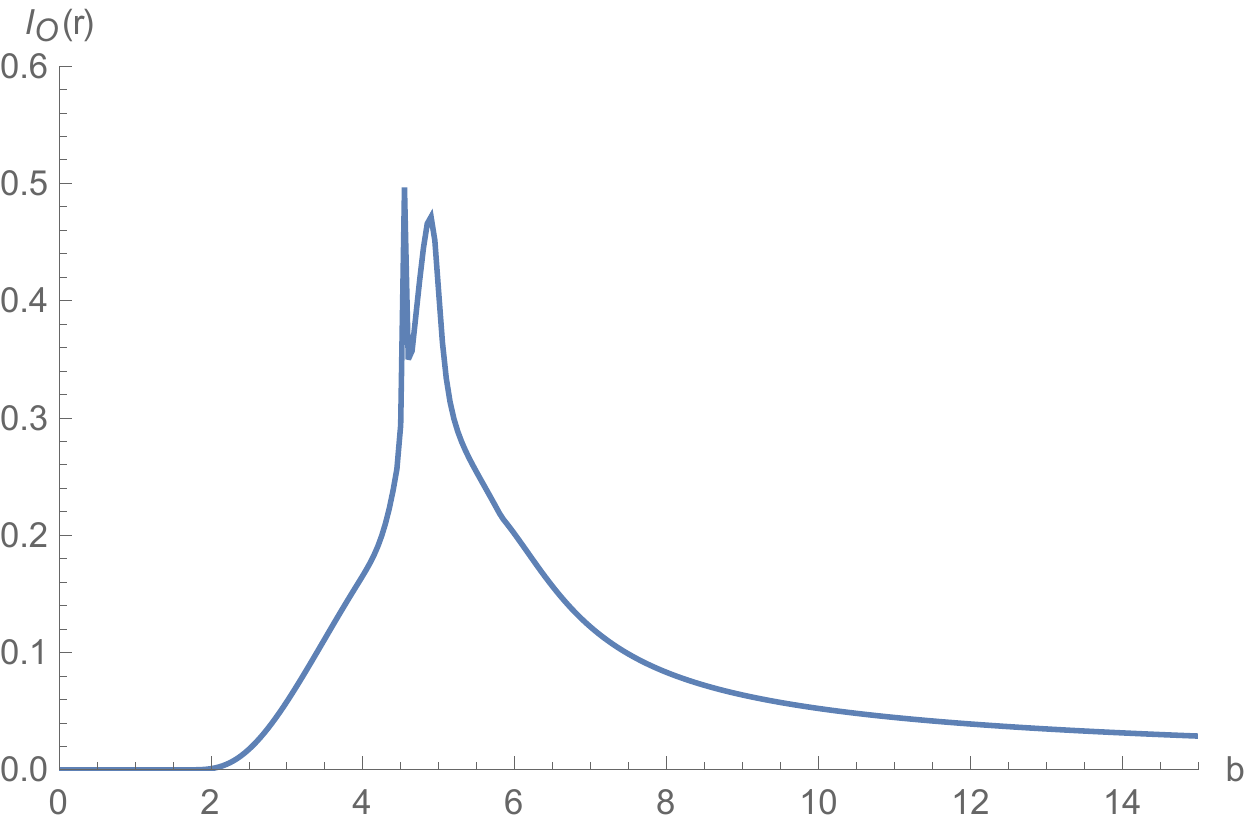}
\includegraphics[width=.28\textwidth]{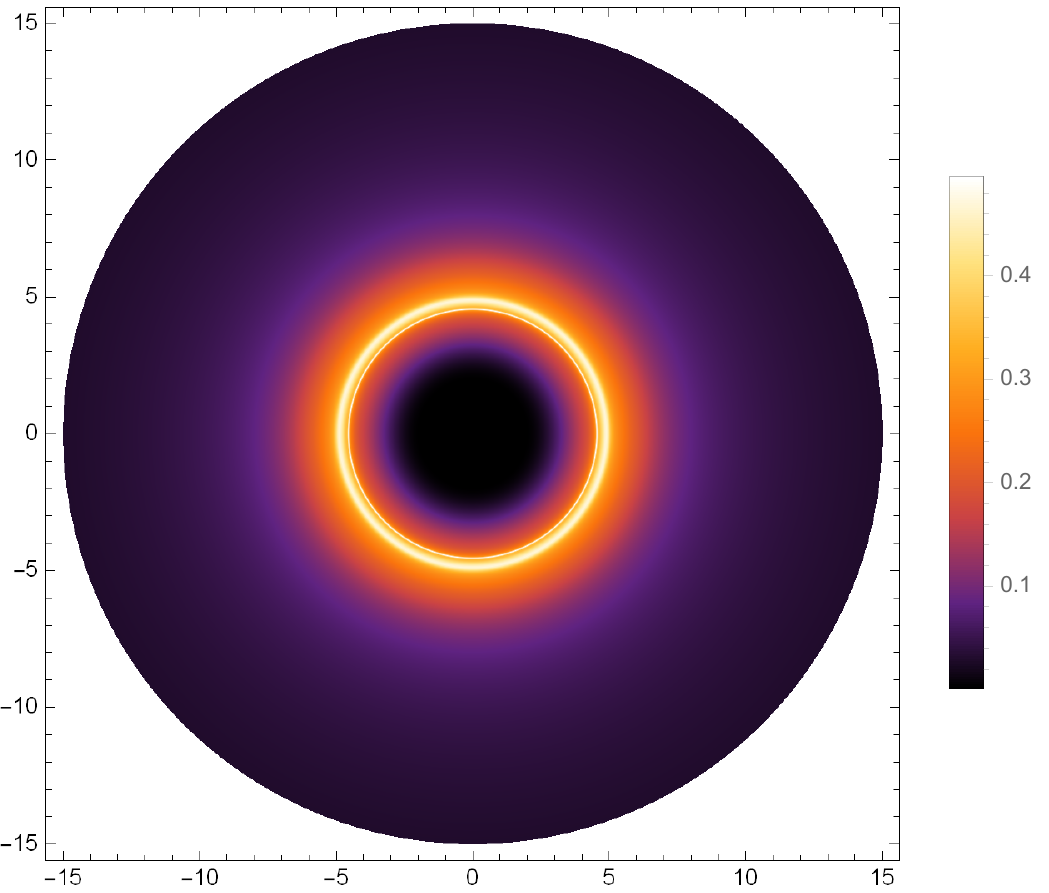}
% "\includegraphics" is very powerful; the graphicx package is already loaded
\caption{\label{fig11}  Observational appearance of the thin disk with different emission profiles  near the black hole, viewed from a face-on orientation. In which, the monopole charge $g=0.75$ and $M=1$.}
\end{figure}

\section{Conclusions and discussions}
\label{conclusion}
It is commonly known that the black hole  is surrounded by the accretion matter, and the distribution of accretion plays a key role in the imaging results of black holes. In the context of Bardeen spacetime,  we have investigated  the  black hole shadow and observation luminosity under the different accretion models. In particular, we pay close attention to the different observation characteristics  of the distant observer when the related physical quantities  is the different values.

When the black hole is surrounded by the spherically symmetric accretion flow, we consider two behaviors of the accretion, i.e., the static and infalling accretion model. For the case of  static spherically symmetric accretions, one can find that the observed luminosity increases with the increase of the monopole charge $g$, while the photon sphere and  radius of black hole decrease. In addition, the maximum observation luminosity always appears in the region of photon sphere under the different values of monopole charge $g$. From the observed intensity density map, one can see that the central region of the shadow is not completely dark,  and the observation brightness of the central region increases slightly with the increase of $g$. In the infalling  accretion model, the maximum observation luminosity is obviously lower than that of the static model, which is the most prominent difference between these two models. The Doppler effect of  infalling matter leads to the difference between the two kinds of spherical accretion. As a result, the observed image in the infalling accretion  model is darker than the static one under the same state parameter. On the other hand,  the change of observation intensity is more obvious  when the monopole charge $g$ takes more larger value. The maximum observation luminosity with $g = 0.75$ is about twice that of  $g = 0$ (Schwarzschild black hole).  It must also be mentioned that in both spherical accretion models, there is a great difference in the observed luminosity, but the position of photon sphere and the shadow radii unchanged under the same parameter, which means that the shadow is an inherent property of spacetime and  not affected by accretion flow behavior.

By considering the Bardeen black hole surrounded by a geometrically and optically thin disk accretion, we have  distinguished the trajectory of light ray  near black hole. Then, we take three different emission models for the thin disk accretion flow. For the observer at infinity ($r_0\rightarrow\infty$), there is not only a dark central  area, but also the photon rings and lensing rings outside of black hole  shadow. In the different emission models, the maximum observation luminosity and the size of   central  dark area are different. The thin disk serves as the only background light source,  the light ray in the region of  photon rings  can accretes through the thin disk more than three times,  and gains enough extra brightness. However,  the photon rings cannot be observed directly for the observer, due to the image of the photon ring region is extremely demagnetized and the photon ring region is very narrow. In addition, the image of lensing ring area with the high demagnetization, leads to the lensing rings can only provide a very small part of the  total observed flux. Therefore, the observation intensity are dominated by  direct emission in these three different emission models.  Although these results are obtained with the ideal models, it is very important to push forward the theoretical study of black hole shadow as far as possible and lay the foundation for the observation results.

%%\appendix
%%\section{Some title}
%%Please always give a title also for appendices.
\vspace{10pt}

\noindent {\bf Acknowledgments}

\noindent
This work is supported  by the National
Natural Science Foundation of China (Grant Nos. 11875095, 11903025), the Sichuan Youth Science and Technology Innovation Research Team(21CXTD0038), and Basic Research Project of Science and Technology Committee of Chongqing (Grant No. cstc2018jcyjA2480).

%%\paragraph{Note added.} This is also a good position for notes added
%%after the paper has been written.


\end{document}